\documentclass[aps,prd,twocolumn,preprintnumbers,balancelastpage,longbibliography,nofootinbib,superscriptaddress]{revtex4-1}

\usepackage{amsmath,amssymb,mathtools,bm}
\usepackage[utf8]{inputenc}
\usepackage{graphicx, color, hepunits}
\usepackage[dvipsnames]{xcolor}
\usepackage{float}
\usepackage[normalem]{ulem}
\usepackage{filecontents}
\usepackage{hyperref} 
\hypersetup{
    colorlinks=true,       
    linkcolor=blue,        
    citecolor=blue,        
    filecolor=magenta,     
    urlcolor=blue          
}
\usepackage[utf8]{inputenc}
\usepackage[english]{babel}

\usepackage{cleveref}
\usepackage[toc,page]{appendix}

\AtBeginDocument{%
    \newwrite\bibnotes
    \def\bibnotesext{Notes.bib}
    \immediate\openout\bibnotes=\jobname\bibnotesext
    \immediate\write\bibnotes{@CONTROL{REVTEX41Control}}
    \immediate\write\bibnotes{@CONTROL{%
    apsrev41Control,author="08",editor="1",pages="1",title="0",year="1"}}
     \if@filesw
     \immediate\write\@auxout{\string\citation{apsrev41Control}}%
    \fi
}%

\newcommand{\order}[1]{\mathcal{O}\left(#1\right)}

\newcommand{\appropto}{\mathrel{\vcenter{
  \offinterlineskip\halign{\hfil$##$\cr
    \propto\cr\noalign{\kern2pt}\sim\cr\noalign{\kern-2pt}}}}}

\definecolor{IBMblue}{rgb}{0.396078431372549, 0.5647058823529412, 0.999999}
\definecolor{IBMyellow}{rgb}{0.996078431372549, 0.6901960784313725, 0.00392156862745098}
\definecolor{IBMmagenta}{rgb}{0.8666666666666667, 0.14901960784313725, 0.5058823529411764}

\begin{document}
\title{Effects of Lighter-than-QCD Axions on Neutron Star Tidal Deformability}
\author{Yonatan Kahn}
\affiliation{Department of Physics, University of Toronto, Toronto, ON, Canada}
\author{Michael Wentzel}\thanks{wentzel4@illinois.edu}
\affiliation{Department of Physics, University of Illinois Urbana-Champaign, Urbana, IL}
\affiliation{Illinois Center for Advanced Studies of the Universe,
University of Illinois Urbana-Champaign, Urbana, IL}
\author{Nicol\'as Yunes}
\affiliation{Department of Physics, University of Illinois Urbana-Champaign, Urbana, IL}
\affiliation{Illinois Center for Advanced Studies of the Universe,
University of Illinois Urbana-Champaign, Urbana, IL}
\preprint{ }

\begin{abstract}
\noindent Finite density corrections to the lighter-than-QCD axion can invert the effective axion potential, sourcing a non-trivial axion field inside dense objects. We perform the first numerical study of the complete dynamics of the lighter-than-QCD axion in a neutron star in 1+1 general relativity, extending the region of analysis to low-mass axions with kilometer-scale Compton wavelengths. We calculate gravitational effects of the axion field on the neutron star and show that for a broad range of axion masses and decay constants, neutron star properties, such as the mass, radius, and compactness, are affected at the ${\cal{O}}(1)$ level. This result indicates that approximate universal $\Lambda-C$ relation for neutron stars is non-trivially broken and can serve as a probe of lighter-than-QCD axions, independent of the unknown nuclear equation of state. We comment on the potential for axion studies with future gravitational-wave observations of neutron stars and applications of this work to other new physics signatures.
\end{abstract}

\maketitle

\section{Introduction}
\label{sec:intro}
\noindent The advent of gravitational-wave (GW) observatories has ushered in a new era of astronomical probes of new physics. The LIGO-Virgo-KAGRA collaboration has observed GWs from neutron star (NS) binary inspirals from dHz to kHz frequencies~\cite{LIGO_NS_1,Abbott_2021}, and advancement of detector, noise, and signal models promise further improvements in sensitivity~\cite{Capote_2025}. Furthermore, near-future space-based and atom interferometers, such as LISA and MAGIS-100, will provide insight into the low frequency GW spectrum from inspiral events on a decade-long time scale~\cite{colpi2024lisadefinitionstudyreport,Abe_2021}. The data from these observatories will probe the macroscopic properties of NSs, allowing for indirect probes of microphysics, including effects of beyond the Standard Model (BSM) physics. In particular, recent work has pointed out that stellar structure is an effective proxy observable of vacuum energy contributions to the nuclear equation of state (EOS)~\cite{Cs_ki_2018,donofrio2025,Panotopoulos_2020}. This effect has been the subject of recent interest in the context of light scalar fields with couplings to the Standard Model (SM) that source the field at sufficiently high SM particle densities~\cite{piinsky,white-dwarf-2024,phiWDs,Zhang_2021,Hook_Huang_2018,Di_Luzio_2021}.

The QCD axion $a$ is one such well-motivated light pseudoscalar field proposed as a solution to the strong CP problem of the SM. This pseudoscalar particle couples to the SM and dynamically cancels the CP violating $\theta$-term at low energies~\cite{PQ1,PQ2,weinberg78,wilczek78,Preskill:1982cy,Abbott:1982af,Dine:1982ah}. In the process, the axion potential is generated dynamically by QCD, leading to an axion mass that depends on the pion mass $m_\pi$ and decay constant $f_\pi$, the axion decay constant $f_a$, and the quark masses $m_u$ and $m_d$: $m_a^{\rm QCD} = [{{m_u m_d}/{(m_u + m_d)^2}}]^{1/2} ({m_{\pi} f_{\pi}}/{f_a})$. However, recent studies of discrete symmetry extensions of the SM symmetry group have shown that a $\mathbb{Z}_N$ symmetry will suppress the mass of an axion that solves the strong CP problem by a factor of $\sqrt{\epsilon} \sim 2^{-N/2 + \log N}$ relative to $m_a^{\rm QCD}$~\cite{Hook_2018}. For such lighter-than-QCD (LQCD) axions, defined by the relation $m_a = \sqrt{\epsilon} \, m_a^{\rm QCD}$ with $0 < \epsilon \leq 1$, finite-density corrections to the axion potential will source a nonzero axion field profile at sufficiently high nucleon densities, in particular inside a NS~\cite{Hook_Huang_2018}.

\begin{figure}[t!]
    \centering
    \includegraphics[width=1.0\linewidth]{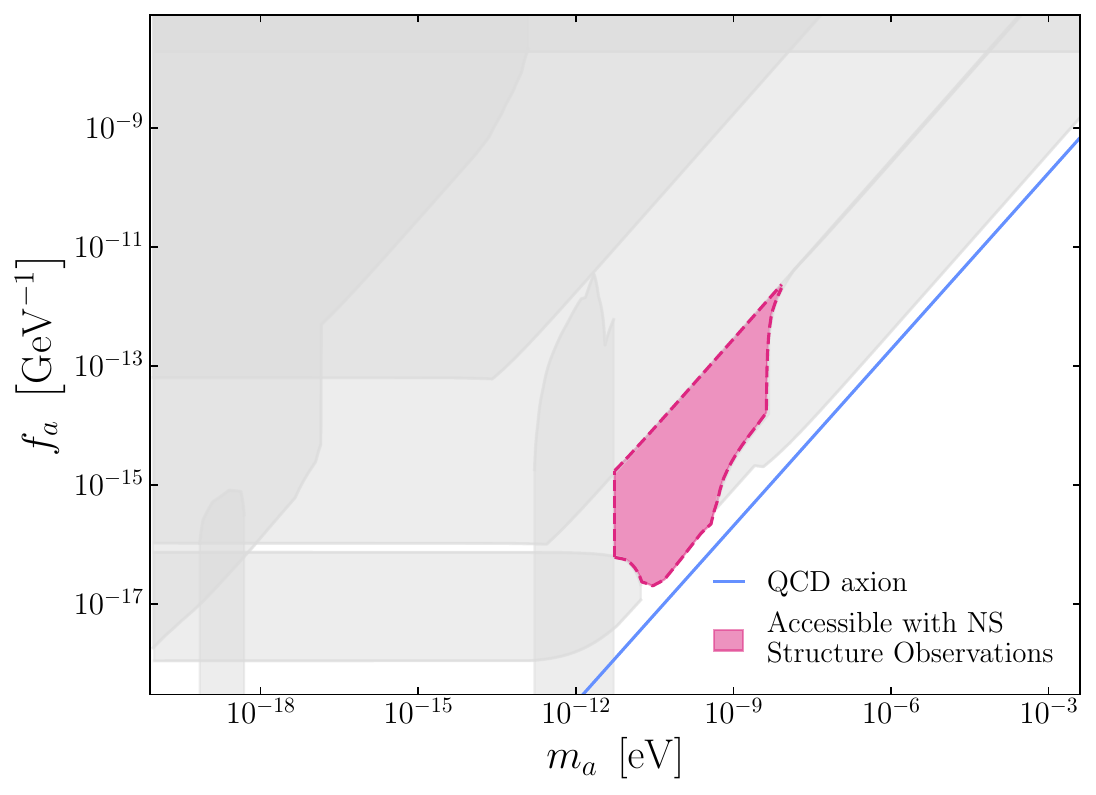}
    \caption{Parameter space of the LQCD axion which highlights the untested region of parameter space (shaded magenta) that is accessible through measurements of NS structure. The QCD axion line, where $m_a = m_a^{\rm QCD} = \beta m_{\pi} f_{\pi}/(2f_a)$  is shown in solid blue. The gray, shaded regions show various constraints from cosmological, astrophysical, and laboratory measurements~\cite{piinsky,white-dwarf-2024,Zhang_2021,Hook_Huang_2018,Baryakhtar_2021,witte2025,Schulthess_2022,Blum_2014,gue2025searchqcdcoupledaxion,Zhang_2023,Karanth_2023,Roussy_2021,Abel_2017,Zhang_2023_2,Alda_2025,banerjee2025oscillatingnuclearchargeradii,Lucente_2022,Caloni_2022,stott2020ultralightbosonicfieldmass,nal_2021,Hoof_2025,Raffelt_2024,iwamoto,Buschmann_2022,Springmann_2025}.}
    \label{fig:LQCD_parameter_space}
\end{figure}

Previous studies have used the effects of the LQCD axion sourced inside a NS on the properties of the host star to constrain significant fractions of the LQCD axion parameter space. In particular, effects of axion fifth forces and power dissipation on GW waveforms~\cite{Hook_Huang_2018,Zhang_2021}, stability of white dwarves in the LQCD axion theory~\cite{white-dwarf-2024}, and over-efficient cooling due to structural changes in NSs~\cite{piinsky} have been used to provide leading constraints on the LQCD axion parameters. Additionally, studies of the cosmological history of the LQCD axion have shown that, for a broad range of axion parameters, the abundance of the LQCD axion exceeds the dark matter abundance~\cite{co2025}. However, a significant region of parameter space, shown in the shaded magenta region in Fig.~\ref{fig:LQCD_parameter_space}, remains viable.

In this paper, we provide the first numerical simulations of the non-linear dynamics of a LQCD axion coupled to a NS in $1+1$ (spherically symmetric, but time dependent and non-linear) general relativity. In particular, previous work has assumed analytic baryon equations of state that do not produce realistic NS observables and/or a static axion field on a NS background without any gravitational backreaction~\cite{piinsky,Hook_Huang_2018}. We here show that a source of damping generically exists in the form of the gravitational self-interactions of the axion field, and we provide numerical simulations of its effect on various NSs and axion parameters. 

Employing the results of our numerical analysis, we study the gravitational effects of the large occupation LQCD axion field sourced by a NS. We show that the radius of the host star can be affected at $\order{1}$, extending the results of previous studies~\cite{piinsky} to axions with characteristic length scales on the order of the NS radius, $m_a^{-1} \gtrsim R_{\rm NS}$. We provide the first calculation of the LQCD axion effect on the NS tidal deformability and find that the approximate universality of the tidal deformability-compactness relation is non-trivially violated at $\order{1}$ in the presence of a LQCD axion. Such an effect is, in principle, observable with current and future gravitational-wave observatories.

The remainder of this paper is organized as follows. In Sec.~\ref{sec:lqcd-axions}, we provide an overview of the LQCD axion. In Sec.~\ref{sec:axion-metric-dynamics}, we derive the coupled equations of motion for nucleonic matter, axion matter, and spacetime metric dynamics. We detail the critical aspects of the numerical simulations in Sec.~\ref{sec:numerical-methods} and discuss common features and useful limiting cases of the dynamics. The gravitational effects of the LQCD axion field on the NS observables are analyzed in Sec.~\ref{sec:observables}. In Sec.~\ref{sec:outlook}, we discuss the prospects and limitations of this work as applied to searches for the LQCD axion and other BSM physics. Additionally, we provide several appendices detailing the more cumbersome and less insightful analytic calculations along with a description of the numerical methods used throughout this paper.

Throughout the paper, we use natural units where $\hbar = c = 1$ and $G_N = 6.7 \times 10^{-39}~\mathrm{GeV}^{-2}$. Useful unit conversions into geometric units, where $G_N = c = 1$, are given in Tab.~\ref{tab:units}.
\begin{table}[t!]
    \centering
    \begin{tabular}{|c|c|c|}
    \hline
    SI & Natural ($\hbar = c = 1$) & Geometric ($\hbar = G_N = 1$) \\
    \hline
    $\hbar$  & 1 & $2.61 \times 10^{-70}~\mathrm{m}^2$ \\
    $G_N$    & $6.7 \times 10^{-39}~\mathrm{GeV}^{-2}$ & 1 \\
    $1~\mathrm{fm}$  & $5.07~\mathrm{GeV}^{-1}$ & $1~\mathrm{fm}$ \\
    $1~\mathrm{J}\cdot\mathrm{m}^{-3}$ & $4.80\times 10^{-38}~\mathrm{GeV}^{4}$ & $8.26\times 10^{-45}~\mathrm{m}^{-2}$ \\
    \hline
    \end{tabular}
    \caption{Useful unit conversions between natural ($\hbar = c = 1$) and geometric ($\hbar = G_N = 1$) units.}
    \label{tab:units}
\end{table}
For readability, we introduce Newton notation for derivatives with respect to coordinate time $t$ and Lagrange notation for derivatives with respect to the radial coordinate $r$. That is, for a function $f(t,r)$, we have
\begin{equation*}
    \frac{\partial f}{\partial t} = \dot{f} \quad \quad \quad \frac{\partial f}{\partial r} = f'~.
\end{equation*}
Derivatives with respect to proper time $\tau$ are always written explicitly, as are derivatives with respect to energy and number densities.

\section{Lighter-than-QCD Axions}\label{sec:lqcd-axions}

The QCD axion $a$ is a well-motivated particle proposed as a solution to the strong CP problem of the SM. This pseudoscalar particle couples to the SM and dynamically cancels the CP violating $\theta$-term at low energies~\cite{PQ1,PQ2,weinberg78,wilczek78,Preskill:1982cy,Abbott:1982af,Dine:1982ah}. In the process, the axion potential $V(a)$ is generated through the anomaly between QCD and the spontaneously-broken Peccei-Quinn symmetry for which the axion is a pseudo-goldstone boson. This produces a low-energy axion potential after QCD confinement of the form
\begin{gather}
    \label{eq:qcd-V}
    V^{\rm QCD}(a) = - m_{\pi}^2 f_{\pi}^2 \left(f(a) - 1 \right) \\
    \label{eq:F-fnc}
    f(a) \equiv \sqrt{1 - \beta^2 \sin^2\left(a / 2 f_a\right)}~,
\end{gather}
for $\beta^2 = {4m_u m_d}/{(m_u + m_d)^2}$. The QCD axion potential leads to an axion dependent pion mass and an axion mass $V''(a)$ fixed in terms of the vacuum pion mass and decay constant: $m_{\pi}^2(a) = m_{\pi}^2 f(a)$, $m_a^{\rm QCD} = \beta {m_{\pi} f_{\pi}}/({2 f_a})$. Additionally, the axion generates corrections to the quark masses $m_q$ that lead to a shift in the nucleon mass $m_N$. Using the fact that $m_{\pi}^2 \propto m_q$, the nucleon mass shift can be parametrized in terms of the nuclear sigma term $\sigma_N \equiv ({\partial m_N}/{\partial m_q}) m_q = ({\partial m_N}/{\partial m_{\pi}^2}) m_{\pi}^2$, which, to leading order in the nucleon number density $n_N$, quantifies how much of the nucleon mass is derived from the quark masses. The nucleon mass shift is\footnote{Additional contributions to the nucleon mass shift from the isovector sigma term $\Delta\sigma_N$ arise at $\order{\Delta\sigma_N / \sigma_N} \sim 0.02$. For a discussion of the associated phenomenology, see Ref.~\cite{piinsky}. In this work, we set $\Delta\sigma_N = 0$, leaving a more detailed accounting to future work.}
\begin{equation}
    \label{eq:mn-shift}
    \begin{aligned}
    \Delta m_N &= \frac{\partial m_N}{\partial m_{\pi}^2} \Delta m_{\pi}^2 + \order{\frac{\sigma_N^2 n_N^2}{m_{\pi}^4 f_{\pi}^4}} \\
    &= \sigma_N \left(f(a) - 1\right) + \order{\frac{\sigma_N^2 n_N^2}{m_{\pi}^4 f_{\pi}^4}}~_.
    \end{aligned}
\end{equation}
To leading order, the effective potential of the QCD axion is the sum of the vacuum potential and the shift in the total rest mass energy density of the nucleons, $\sigma_N n_N (f(a) - 1)$:
\begin{equation}
    \label{eq:qcd-Veff}
    V^{\rm QCD}_{\rm eff}(a) = - m_{\pi}^2 f_{\pi}^2\left(1 - \frac{\sigma_N n_N}{m_{\pi}^2 f_{\pi}^2} \right)\left(f(a) - 1\right)~_.
\end{equation}
Since $f(a) < 1$, the effective QCD axion potential is minimized when $f(a)$ is maximized, which occurs at $a = 0$ in vacuum. 

In sufficiently dense nucleon environments with $n_N > m_{\pi}^2 f_{\pi}^2 / \sigma_N$, the potential could become inverted, leading to a minimum at $a = \pi f_a$, but the domain of validity of Eq.~(\ref{eq:qcd-Veff}) is limited by the expansion in $\sigma_N^2 n_N^2 / m_{\pi}^4 f_{\pi}^4$ to $n_N < m_{\pi}^2 f_{\pi}^2 / \sigma_N$. However, recent studies of discrete symmetry extensions of the SM symmetry group have shown that a $\mathbb{Z}_N$ symmetry will suppress the mass of an axion that solves the strong CP problem by a factor of $\sqrt{\epsilon} \sim 2^{-N/2 + \log N}$ relative to $m_a^{\rm QCD}$~\cite{Hook_2018}. For such light QCD (LQCD) axions, defined by the relation
\begin{equation}
    \label{eq:lqcda-mass}
    m_a = \sqrt{\epsilon} \, m_a^{\rm QCD}
\end{equation}
with $0 < \epsilon \leq 1$, the zero density term of the potential in Eq.~(\ref{eq:qcd-Veff}) is suppressed by a factor of $\epsilon$.
\begin{equation}
    \label{eq:lqcd-Veff}
    V^{\rm LQCD}_{\rm eff}(a) = - m_{\pi}^2 f_{\pi}^2\left(\epsilon - \frac{\sigma_N n_N}{m_{\pi}^2 f_{\pi}^2} \right)\left(f(a) - 1\right)
\end{equation}

\begin{figure}[t!]
    \centering
    \includegraphics[width=1.0\linewidth]{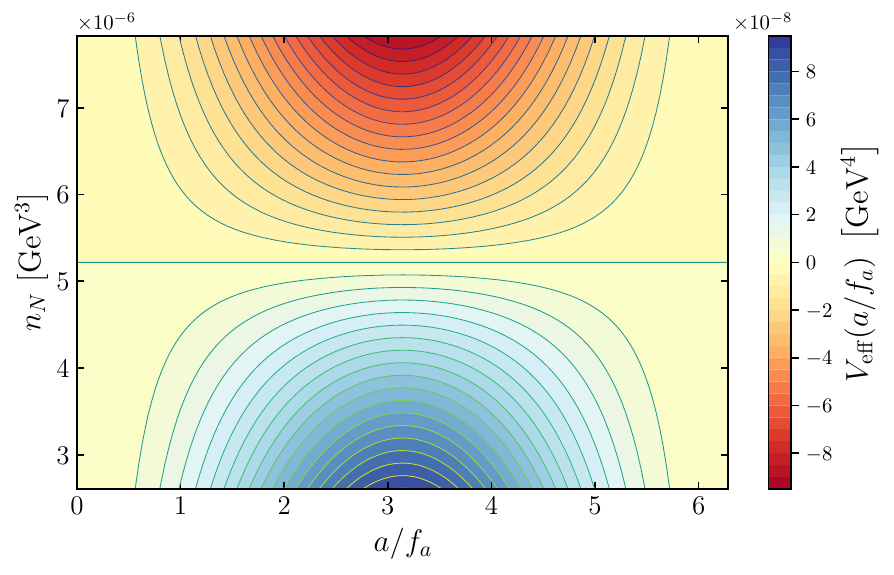}
    \caption{Contours for the effective axion potential $V_{\rm eff}^{\rm LQCD}(a)$ plotted for $\epsilon = 10^{-3}$. For this value of $\epsilon$, the critical nucleon number density is $n_N^{\rm crit} = 5.2 \times 10^{-6}~\mathrm{GeV}^3$. Negative values of the effective potential correspond to the suppression of the nucleon mass, and at zero nucleon number density, $V_{\rm eff}^{\rm LQCD}$ is strictly positive.}
    \label{fig:potential-contours}
\end{figure}

For LQCD axions at zero nucleon density, $n_N = 0$, the minimum of the potential is still achieved for $a = 0$. However, the LQCD axion potential can change sign even when $n_N < m_{\pi}^2 f_{\pi}^2 / \sigma_N$, in which case Eq.~(\ref{eq:qcd-Veff}) is still under perturbative control. Thus, finite-density corrections to the axion potential will source a nonzero axion field profile $a \sim \pi f_a$ at sufficiently high nucleon densities. For a given $\epsilon$, the condition on the number density to source an axion field is determined by $(d V^{\rm LQCD}_{\rm eff}(a_0) / da) = 0$, where
\begin{equation}
    \label{eq:axion-sourcing-condition}
  \frac{a_0}{f_a}  = \left\{ \begin{array}{ll}
        0, & n_N < n_N^{\rm crit} \\ 
        \pi, & n_N > n_N^{\rm crit}.\\
    \end{array}\right.
\end{equation}
and $n_N^{\rm crit} = \epsilon \, m_{\pi}^2 f_{\pi}^2 / \sigma_N$ is the critical number density for the axion field to be minimized at $a = \pi f_a$ instead of $a = 0$. The potential contours as a function $a / f_a$ and $n_N$ are shown in Fig.~(\ref{fig:potential-contours}) for $\epsilon = 10^{-3}$, giving a critical density of $n_N^{\rm crit} \simeq 5 \times 10^{-6}~\mathrm{GeV}^3$. Neutron stars have typical central number densities that exceed nuclear saturation density $n_{\rm sat} \simeq 1.2 \times 10^{-3}~\mathrm{GeV}^3$, and therefore, they can source a nonzero axion field~\cite{Hook_Huang_2018}. 

Previous studies have examined several signatures of non-trivial LQCD axion profiles in NSs, producing constraints across a significant fraction of the LQCD parameter space, as shown in Fig.~\ref{fig:LQCD_parameter_space}. Fifth force effects from the axion profile have been shown to cause power dissipation during NS inspirals, affecting the GW waveform of these events and constraining the LQCD parameter space for $f_a \gtrsim 10^{16}~\mathrm{GeV}$~\cite{Hook_Huang_2018}. Earth-based measurements of the pion mass and observation of efficient nuclear chain reactions in the Sun constrain $\epsilon \lesssim 10^{-15}$ for $f_a \lesssim 10^{15}~\mathrm{GeV}$~\cite{Hook_Huang_2018}. For $\epsilon \lesssim 10^{-7}$ and $f_a\lesssim 10^{16}~\mathrm{GeV}$, it has been shown that the LQCD axion will destabilize white dwarf stars (WD), and limits on NS cooling provide strong constraints on $\epsilon \lesssim 0.1$ and $f_a \lesssim 10^{14}~\mathrm{GeV}$~\cite{piinsky}. Complementary constraints also exist from the superradiant instability of stellar-mass black holes~\cite{witte2025,Baryakhtar_2021} and various laboratory and cosmological constraints~\cite{white-dwarf-2024,Zhang_2021,Schulthess_2022,Blum_2014,gue2025searchqcdcoupledaxion,Zhang_2023,Karanth_2023,Roussy_2021,Abel_2017,Zhang_2023_2,Alda_2025,banerjee2025oscillatingnuclearchargeradii,Lucente_2022,Caloni_2022,stott2020ultralightbosonicfieldmass,nal_2021,Hoof_2025,Raffelt_2024,iwamoto,Buschmann_2022,Springmann_2025}.

The remaining region of parameter space, shown in shaded magenta in Fig.~\ref{fig:LQCD_parameter_space}, has proven difficult to rule out due to the complex, nonlinear dynamics of the axion and nucleonic matter, along with the large uncertainties in the EOS for dense nuclear matter that governs NS observables. Recent analysis of the cosmological dynamics of the LQCD axion argues that, in this region of parameter space, for an $\order{1}$ axion misalignment angle, solving the strong CP problem and producing a fraction of dark matter $f_{\rm DM} \leq 1$ is improbable~\cite{co2025}, but the argument does not directly rule out the LQCD axion. 

In this work, we demonstrate how this parameter space may be probed through a numerical treatment of the full dynamics of an LQCD axion and a NS system. As we will show in the following sections, this fully dynamical treatment shows that the tidal deformability -- compactness relation, an approximately EOS-invariant relation~\cite{Yagi_2013,Yagi_2013_2,Nath_2023}, is affected at an $\order{1}$ level by the LQCD axion.

\section{Dynamics of Axion-Baryon Stars}\label{sec:axion-metric-dynamics}
As described in the previous section, finite nucleon density corrections to the axion potential can manifest inside a NS for LQCD axions, resulting in an axion field with a potential minimum at $a = \pi f_a$ rather than the standard vacuum minimum at $a = 0$. The non-zero energy density and pressure of the resultant axion field can impact the distribution of nucleonic matter within the host NS and affect the macroscopic gravitational observables, such as its mass, radius, and tidal deformability. In the region of LQCD axion parameter space where $10^{-6} \lesssim \epsilon \lesssim 10^{-1}$ and $10^{13}~\mathrm{GeV} \lesssim f_a \lesssim 10^{16}~\mathrm{GeV}$, the Compton wavelength of the axion is $\order{\mathrm{km}}$, giving rise to a non-trivial axion profile inside the NS and a field profile outside the star with spatial extent comparable to the NS radius, necessitating a simultaneous treatment of the axion and NS dynamics. This is the generalization of the large axion mass scenario in which the axion profile can be approximated as a step function given that the width of the domain wall of the axion field is much smaller than the radius of the NS. In this section, we derive the coupled equations of motion for the axion matter, nucleonic matter, and metric.

The complete action for the system is written in terms of metric determinant $g$ and the sum of the nucleon, axion, and Einstein-Hilbert Lagrangians:
\begin{equation}
    \label{eq:action}
    S_{\rm tot} = \int d^4x \sqrt{-g} \left(\mathcal{L}_{N} + \mathcal{L}_a + \mathcal{L}_{\rm EH}\right).
\end{equation}
The nucleon Lagrangian is proportional the nucleonic energy density $\rho_N$~\cite{Hawking_Ellis_1973} 
\begin{equation}
    \label{eq:LB}
    \mathcal{L}_{N} = - \rho_N~,
\end{equation}
the axion Lagrangian is given in terms of the effective potential of Eq.~(\ref{eq:lqcd-Veff})
\begin{equation}
    \label{eq:La}
    \mathcal{L}_a = \frac{1}{2} \partial^{\mu} a \partial_{\mu} a - V_{\rm eff}^{\rm LQCD}(a)~,
\end{equation}
and the Einstein-Hilbert Lagrangian is proportional to the Ricci scalar $R$
\begin{equation}
    \label{eq:LEH}
    \mathcal{L}_{\rm EH} = \frac{1}{16 \pi G_N} R~.
\end{equation}
The coupled equations of motion (EOMs) for the system are derived from Eq.~(\ref{eq:action}) by calculating the stress-energy tensor from the variation of the action with respect to the metric and applying the Euler-Lagrange equations. The resultant EOMs will be the Einstein equations, stress-energy conservation, and a modified Klein-Gordon equation for the axion field.

We choose to work under the simplifying assumption of a spherically-symmetric system, such that the metric can be parametrized in terms of a $(t,t)$ component and an $(r,r)$ component. In a $(+,-,-,-)$ signature, the line element can thus be written as~\cite{MTW}
\begin{equation}
    \label{eq:metric}
    ds^2 = F(t,r)^2 dt^2 - H(t,r)^2 dr^2 - r^2 d\theta^2 - r^2 \sin^2\theta d\phi^2~_.
\end{equation}
Additionally, we require that the baryon 4-velocity, $u^{\mu} = d x^{\mu} / d\tau$ for proper time $\tau$, be timelike, which in the mostly-minus normalization implies
\begin{equation}
    \label{eq:velocity-constraint}
    u^{\mu} u_{\mu} = 1~.
\end{equation}
With the assumption of spherical symmetry, there are only two non-zero components of the 4-velocity, $u^t$ and $u^r$. Applying the normalization in Eq.~(\ref{eq:velocity-constraint}) and requiring that $u^t = dt/d\tau > 0$, we can define the 4-velocity in terms of a single, scalar function $u(t,r)$ as
\begin{equation}
    \label{eq:4-vel}
    u^{\mu} = \frac{1}{\sqrt{1 - u^2}} \left( \frac{1}{F}, \frac{u}{H}, 0, 0 \right).
\end{equation}

\subsection{Stress-Energy Tensor}
The symmetric and gauge-invariant Hilbert stress-energy tensor is defined by the variation of the action with respect to the metric, normalized by the metric determinant~\cite{Hawking_Ellis_1973}:
\begin{equation}
    \label{eq:EH-SET}
    T_{\mu\nu}^{(i)} = \frac{2}{\sqrt{-g}} \frac{\delta S_{(i)}}{\delta g^{\mu\nu}} = -g_{\mu\nu}\mathcal{L}_{(i)} + 2\frac{\delta \mathcal{L}_{(i)}}{\delta g^{\mu\nu}}~,
\end{equation}
for $i \in \{N, a, \mathrm{EH}\}$. The calculation is straightforward; we will derive the SET for the baryon fluid and the axion separately. 

\begin{figure*}[t!]
    \centering
    \includegraphics[width=1.0\linewidth]{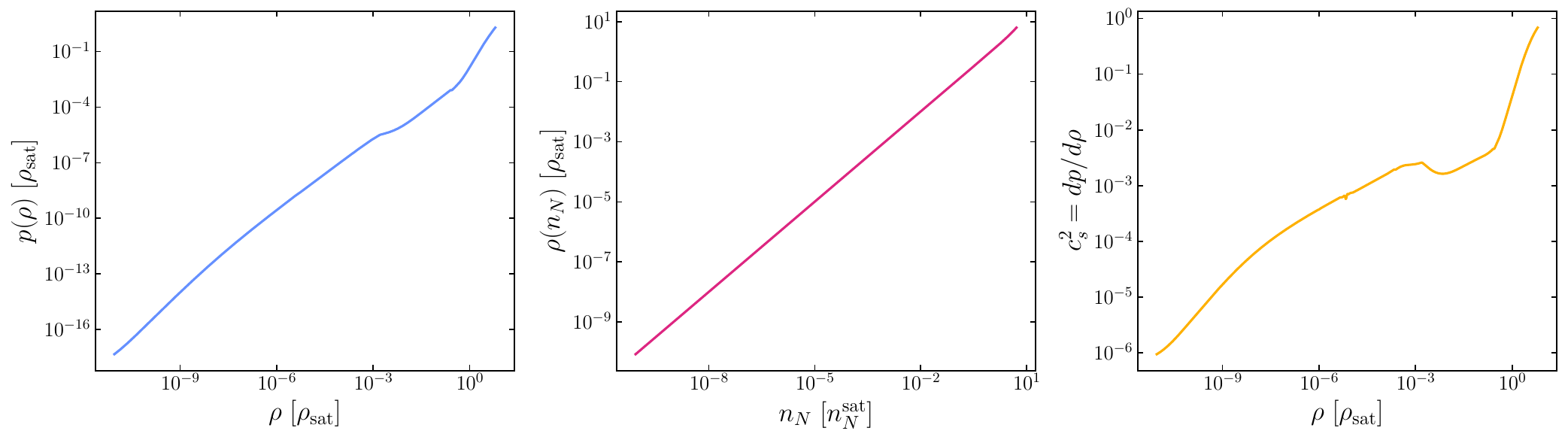}
    \caption{The EOS $p(\rho)$ (LEFT), the nucleon energy density as a function of nucleon number density $\rho(n_N)$ (CENTER), and the speed of sound $c_s^2 = dp / d\rho$ (RIGHT) for the SLy4 EOS. Observe that the EOS remains causal (i.e.~$c_s^2 < 1$) for the range of densities considered in this work.}
    \label{fig:EOS-info}
\end{figure*}

The baryon SET, derived from Eq.~(\ref{eq:EH-SET}) with $i = N$ in Appendix~\ref{app:nucleon-SET}, is simply that of a perfect fluid
\begin{equation}
    \label{eq:NS-set}
    T_{\mu\nu}^{(N)} = (\rho_N + p_N) u_{\mu} u_{\nu} - p_N g_{\mu\nu}
\end{equation}
where the pressure $p_N$ is given in terms of the EOS $p_N(\rho_N)$, the energy density $\rho_N$ is given in terms of the number density as $\rho_N(n_N)$. Throughout this work, we use the zero temperature, $\beta$-equilibrium SLy4 EOS. The SLy4 EOS utilizes the Skyrme effective potential for nucleons to model the interactions high-density nuclear matter, producing NS phenomenology consistent with observations~\cite{compose,Typel2013,Oertel,Chabanat1997,sly4,Danielewicz_2009}. In Fig.~(\ref{fig:EOS-info}), we show the nucleon EOS $p_N(\rho_N)$ and the speed of sound $c_s^2 = dp_N / d\rho_N$ both as a function of energy density, and the energy density as a function of number density $\rho_N(n_N)$ for the SLy4 EOS. The true EOS for dense nuclear matter is unknown, but the SLy4 EOS produces macroscopic NS observables that are consistent with the limited observational data presently available. In general, some of the results presented here will, of course, depend on the EOS, although we expect that the trends we observe as we vary the axion potential parameters $\epsilon$ and $f_a$ are preserved for other EOSs. We will indicate throughout the text where we expect the results to be qualitatively similar for different EOSs, and we will calculate an EOS-insensitive observable in Sec.~\ref{sec:observables}.

The axion SET is calculated from Eq.~(\ref{eq:EH-SET}) with $i = a$:
\begin{equation}
    \label{eq:axion-set}
    \begin{aligned}
    T^{(a)}_{\mu\nu} = &\partial_{\mu} a \partial_{\nu} a - \frac{1}{2} g_{\mu\nu} \partial^{\alpha} a \partial_{\alpha} a  \\ 
    &+ g_{\mu\nu} V_{\rm eff} + n_N \frac{d V_{\rm eff}}{d n_N} \left(u_{\mu}u_{\nu} - g_{\mu\nu}\right)~_,
    \end{aligned}
\end{equation}
where the final term arises from the fact that the variation of the number density with the metric is $\delta n_N / \delta g^{\mu\nu}~=~-(1/2)~n_N~(u_{\mu}~u_{\nu}~-~g_{\mu\nu})$ ( see Appendix \ref{app:nucleon-SET} and Ref.~\cite{Hawking_Ellis_1973}). 

The energy density $\rho_a = T^{(a) t}_{\quad \ t}$, radial pressure $p_a^{(r)} = T^{(a) r}_{\quad \ r}$, angular pressure $p_a^{(\Omega)} = T^{(a) \theta}_{\quad \ \theta} = T^{(a) \phi}_{\quad \ \phi}$, and the radial energy flux $\varphi_a^{(r)} = T^{(a) t}_{\quad \ r}$ of the axion field are respectively
\begin{gather}
    \label{eq:axion-energy-density}
    \rho_a = \frac{\dot{a}^2}{2F^2} + \frac{a'^2}{2H^2} + (1 - f(a)) \left(\epsilon m_{\pi}^2 f_{\pi}^2 + \frac{\sigma_N n_N}{u^2 - 1}\right), \\
    \label{eq:axion-radial-pressure}
    p^{(r)}_a = \frac{\dot{a}^2}{2F^2} + \frac{a'^2}{2H^2} - (1 - f(a)) \left(\epsilon m_{\pi}^2 f_{\pi}^2 + \frac{\sigma_N n_N u}{u^2 - 1}\right), \\
    \label{eq:axion-angular-pressure}
    p^{(\Omega)}_a = \frac{\dot{a}^2}{2F^2} - \frac{a'^2}{2H^2} - (1 - f(a)) \epsilon m_{\pi}^2 f_{\pi}^2, \\
    \label{eq:axion-energy-flux}
    \varphi^{(r)}_a = - \frac{\dot{a}a'}{H^2} + (1 - f(a))\sigma_N n_N \frac{uF}{H(u^2 - 1)}~.
\end{gather}
Note that, unlike the baryonic perfect fluid, where $p_N^{(r)} = p_N^{(\Omega)}$, the axion introduces a non-isotropic pressure component, a feature crucial to the calculation of the tidal deformability. Furthermore, in the static limit, where $u \rightarrow 0$, the finite density corrections to the LQCD axion potential affect only the energy density, not the pressure. This fact can be understood intuitively by noting that the finite density piece of the potential arises as a correction to the nucleon mass (see Eq.~(\ref{eq:mn-shift})). Since the nucleonic pressure depends on the internal energy of the nucleons and not on the mass density, this correction should not affect the total pressure. 

The form of the radial pressure suggests that, since $V(a) \geq 0$, the axion pressure can become negative. In fact, for sufficiently small nucleonic pressure, it is possible that $p_N(\rho_N) + p_a^{(r)} < 0$ at densities $\rho_N(n_N)$, where $n_N > n_{\rm crit}$, where recalled that $n_{\rm crit}$ was defined below Eq.~(\ref{eq:axion-sourcing-condition}). That is, it is possible to source an axion field in a region where the axion pressure dominates over the nucleonic pressure, leading to a negative total pressure. This feature of the LQCD axion, first noted for white dwarfs in Ref.~\cite{white-dwarf-2024} and extended to NS in Ref.~\cite{piinsky}, indicates that the effect of the axion cannot be considered perturbatively. In general, the negative pressure in the axion field will lead to intricate NS dynamics, as the axion falls toward the NS and is compressed inwards, while simultaneously being sourced by the nucleon number density in the negative pressure region. A reasonable expectation for the steady-state solution of such a system is a nucleon number density that sharply drops to zero, such that no negative pressure region exists. Such solutions were shown to exist for the static system in the infinitely-thin axion domain wall approximation in Ref.~\cite{piinsky} (NS) and Ref.~\cite{white-dwarf-2024} (WD). In each case, the axion leads to stellar density profiles without crust layers, modifying the thermal loss and structure of the star. To our knowledge, however, the coupled LQCD axion-NS system has not yet been analyzed in the case of a non-trivial axion profile or in a non-static system.

\subsection{Equations of Motion}

The EOMs are derived straightforwardly from Eq.~(\ref{eq:action}), and one obtains the Einstein equations, the stress-energy consideration equation, and the axion equation of motion, namely
\begin{gather}
    \label{eq:EEs}
    G_{\mu\nu} = 8 \pi G_N \left(T^{(N)}_{\mu\nu} + T^{(a)}_{\mu\nu}\right),\\
    \label{eq:divergence-constraint}
    \nabla_{\mu} \left( T_{(N)}^{\mu\nu} + T_{(a)}^{\mu\nu} \right) = 0, \\
    \label{eq:axion-EOM}
    \Box a = - \frac{d V_{\rm eff}}{da}~_,
\end{gather}
where $\Box = g^{\mu\nu} \nabla_{\mu} \partial_{\nu}$ is the D'Alembertian operator acting on a scalar in a curved spacetime metric. 

After applying the 4-velocity constraint and the nucleon EOS $p_N(\rho_N)$, the system of equations has five unknown functions: the metric coefficients $F$ and $H$, the nucleon number density $n_N$ and the 4-velocity function $u$, and the axion field $a$. The Einstein equations yield three independent equations from the $(t,t)$, $(r,r)$, and $(t,r)$ components of Eq.~(\ref{eq:EEs}). The resultant equations can be expressed as
\begin{gather}
    \label{eq:dFdr}
    \frac{\partial F}{\partial r} = \frac{\left[8 \pi r^2\, \, p^{(r)} + 1\right]H^2 F - F}{2r}\\
    \label{eq:dHdr}
    \frac{\partial H}{\partial r} = \frac{\left[8 \pi r^2 G_N \rho - 1\right] H^3 + H}{2r}\\
    \label{eq:dHdt}
    \begin{aligned}
    \frac{\partial H}{\partial t} = 4 \pi r G_N H^3 \varphi^{(r)}~,
    \end{aligned}
\end{gather}
where $\rho$ and $p^{(r)}$ are the total density and total radial pressure:
\begin{gather}
    \label{eq:total-density}
    \rho = \rho_N + \rho_a \\
    \label{eq:total-radial-pressure}
    p^{(r)} = p_N + p_a^{(r)}~.
\end{gather}
The total angular pressure and the total radial energy flux can be defined respectively as well:
\begin{gather}
    \label{eq:total-angular-pressure}
    p^{(\Omega)} = p_N + p^{(\Omega)}_a \\
    \label{eq:total-radial-energy-flux}
    \varphi^{(r)} = \varphi^{(r)}_N + \varphi^{(r)}_a~.
\end{gather}

Equations~(\ref{eq:dFdr})--(\ref{eq:dHdr}) should be understood as the standard Tolman-Oppenheimer-Volkoff (TOV) equations~\cite{tolman39,oppenheimer39} modified by an axion energy density and pressure.\footnote{Here, we have expressed the TOV equations directly in terms of the metric coefficients $F$ and $H$, rather than using the parametrization $\nu = 2 \log F$ and $\lambda = 2 \log H$, commonly employed in the literature.} Setting $\rho = \rho_N$ and $p^{(r)} = p_N$ reproduces the TOV equations and can be used to solve for the static macroscopic properties of NS in the absence of a LQCD axion field, such as the mass, radius, and tidal deformability of a star. We will discuss this procedure in more detail later, when evaluating the initial state of the dynamic axion-baryon star simulation. Equation~(\ref{eq:dHdt}) highlights that the temporal evolution of $H$ is proportional to the radial energy flux in the star, and can be used to eliminate the dependence in the matter equations of motion on metric derivatives. 

The stress-energy conservation equation [Eq.~(\ref{eq:divergence-constraint})] yields evolution equations for the baryon fields $n_N$ and $u$. In terms of the axion density, pressures, and energy flux, the time derivatives of the baryon fields are
\begin{widetext}
    \begin{equation}
    \label{eq:dnNdt}
    \begin{aligned}
        \frac{dn_N}{dt} = &\Bigg\{2 H^2 r u \dot{F} \varphi_a^r+F H r \left[\varphi _a^r \left(\left\{u^2+1\right\} F'-6 u \dot{H}\right)-2 H u \dot{\varphi _a^r}\right] +H F^2 \left(u^2+1\right) \left[2 \varphi _a^r+r \left(\dot{\rho _a}+\left\{\varphi _a^r\right\}'\right)\right] \\
        & \quad \quad + r F^2 \left[H' \varphi _a^r+p \dot{H}+\dot{H} \left(\left\{u^2+1\right\} p_a^{(r)}+\rho -u^2 \left\{p+\rho -\rho _a\right\}+\rho _a\right)+u \left(u \varphi _a^r H'-2 \left\{p_a^{(r)}+\rho _a\right\} F'\right)\right] \\
        & \quad \quad +F^3 \left[r (p+\rho ) u'+u \left(-4 p_a^{(r)}+4 p_a^{(\Omega)}+2 p+2 \rho -2 r \left\{p_a^{(r)}\right\}'-\left\{\frac{dp}{d\rho}-1\right\} r n_N' \rho '\left(n_N\right)\right)\right]\Bigg\}, \\
        &\times \Bigg\{F^2 H r \left[\frac{dp}{d\rho} u^2-1\right] \rho '\left(n_N\right)\Bigg\}
    \end{aligned}
    \end{equation}
    \begin{equation}
    \label{eq:dudt}
    \begin{aligned}
        \frac{du}{dt} = &\Bigg\{H^2 r \left(u^2-1\right) \left[\frac{dp}{d\rho} u^2+1\right] \dot{F} \varphi _a^r -F^3 \left[\left(\frac{dp}{d\rho}-1\right) r u (p+\rho ) u'\right] -ru \left[p_a^{(r)}+p+\rho +\rho _a\right] F' \\
        &\quad\quad  -F H r \left(u^2-1\right) \left[\left(3 \left\{\frac{dp}{d\rho} u^2+1\right\} \dot{H}-\left\{\frac{dp}{d\rho}+1\right\} u F'\right) \varphi _a^r+H \left(\frac{dp}{d\rho} u^2+1\right) \dot{\varphi _a^r}\right] \\
        &\quad\quad  +F^2 \left(u^2-1\right) \left[ r u H' \varphi _a^r+ r u\dot{H} \left(p_a^{(r)}-p-\rho +\rho _a+\frac{dp}{d\rho} \left\{p_a^{(r)}+p+\rho +\rho _a\right\}\right)\right] \\
        &\quad\quad   + F^2 \left(u^2-1\right) \left( r u \frac{dp}{d\rho} \left(H' \varphi _a^r+u \left(-p_a^{(r)}+p+\rho -\rho _a\right) F'\right) +(\frac{dp}{d\rho}+1) H \left(2 \varphi _a^r+r \left(\dot{\rho _a}+\left(\varphi _a^r\right)'\right)\right)\right) \\
        &\quad\quad  - F^3 \left(u^2-1\right) \left[2\left(\frac{dp}{d\rho} u^2+1\right) \left(p_a^{(r)} - p_a^{(\Omega)} + \frac{r}{2} p_a^{r'}\right) +\frac{dp}{d\rho} \left\{-2 (p+\rho ) u^2-r \left(u^2-1\right)\rho'\right\}\right]\Bigg\} \\
        &\times \Bigg\{F^2 H r \left[\frac{dp}{d\rho} u^2-1\right] (p+\rho )\Bigg\}.
    \end{aligned}
    \end{equation}
\end{widetext}
Again, setting $\rho = \rho_N$, $p^{(r, \Omega)} = p_N$, and $\varphi^{(r)} = \varphi^{(r)}_N$ reduces the evolution equations to those of a purely baryonic NS. Equations~(\ref{eq:dnNdt})--(\ref{eq:dudt}) clearly demonstrate that the axion field contributes to the evolution of the baryon fields both indirectly (through its impact on the metric coefficients) but also explicitly (through $\rho_a$, $p_a^{(r,\Omega)}$, and $\varphi^{(r)}_a$). The latter results from the interaction term in Eq.~(\ref{eq:lqcd-Veff}) and highlights the non-perturbative nature of the system. In other words, the effect of the axion on the baryon fields \textit{cannot} be modeled through axion metric perturbations acting on the baryon fields; there are additional physics that can only be probed in the coupled, nonlinear system.

Finally, the axion EOM in Eq.~(\ref{eq:axion-EOM}) can be written explicitly as
\begin{widetext}
    \begin{equation}
        \label{eq:axion-EOM-explicit}
        \frac{\ddot{a}}{F^2} - \frac{a''}{H^2} + \frac{\dot{a}}{F^2} \frac{\partial}{\partial t}\log \frac{H}{F} + \frac{a'}{H^2} \frac{\partial}{\partial r}\log \frac{H}{r^2 F} = -\frac{\beta^2 \sin^2\left(\frac{a}{f_a}\right)\left(\epsilon m_{\pi}^2 f_{\pi}^2 - \sigma_N n_N\right)}{4 f_a f(a)}.
    \end{equation}
\end{widetext}
The qualitative dynamics of the axion can be inferred from the above equation: the axion equation of motion is a damped wave equation in curved space. In this form, it is clear that a frictional term exists in the equation of motion, $F^{-2} \frac{\partial}{\partial t}\log H/F$. As the axion field oscillates about the potential minimum, $H/F$ will oscillate due to the changing contribution to the energy density and pressure from the axion field. Therefore, the frictional term will, in fact, oscillate between a friction force and an anti-friction force. Inside the star, the behavior of the friction term is highly non-linear. However, we will verify in Sec.~\ref{sec:evolution-and-steady-state} that the term proportional to $\dot{a}$ behaves dominantly as a damping term, resulting in the rapid decay of axion dynamics inside the NS. Conversely, outside the star, when the axion field is no longer sourced, the metric components decay to flat spacetime, such that $F\sim 1 \sim H$; thus, outside the star, the damping/anti-damping goes to zero. In Sec.~\ref{sec:evolution-and-steady-state}, we show that the decay rate of the axion oscillations outside the star is significantly smaller than the decay rate of internal oscillations. 

\subsection{Additional Sources of Loss}

In addition to the damping which naturally arises from the coupling of the axion to the metric in Eq.~(\ref{eq:axion-EOM-explicit}), other loss channels are expected to exist in the axion-NS system, in particular from the finite quality factor of NS normal modes~\cite{roy2024analysisneutronstarfmode,mohanty2024astrophysicalconstraintsneutronstar,QNM,Kokkotas_1999,gomez}. In general, motion of the axion field will couple to the nucleonic matter fields, the excitations of which can be decomposed into normal modes. Each of these modes will have a finite quality factor resulting in the relaxation of any mode excitations through thermal loss. Practically, the quality factors are strongly dependent on the nuclear EOS and are outside the scope of this work. It is sufficient to note that current estimates of NS normal mode quality factors are $Q_i \sim \order{100}$~\cite{roy2024analysisneutronstarfmode}. We therefore expect oscillations at frequencies $\omega \gtrsim \omega_{\rm min} \sim 1~\mathrm{Hz}$ --- the minimum resolvable frequency in our simulations given the total physical simulation time and a frequency smaller than the smallest axion mass we consider in this work --- to decay within $\tau_i \lesssim 10^{2}~\mathrm{s}$. This is well below the age of the youngest observed NS, indicating that any NS with an added axion field will have reached a steady state long before observation. As such, there is no observable consequence of the added damping. With this in mind, we implement a damping term of the following form.
\begin{equation}
    \label{eq:NS-normal-damping}
    \frac{\partial u}{ \partial t} \rightarrow \frac{\partial u}{\partial t} - \alpha \frac{u}{\Delta r}
\end{equation} 
We verify that the results of the simulations are insensitive to the choice of $\alpha$ in Appendix~\ref{app:numerics}; in the simulations presented below, we set $\alpha = 10^{-3}$.

\subsection{Limiting Cases and Static Solutions}

Previous analyses of the LQCD axion inside a NS have considered only fully static solutions for $F$, $H$, $n_N$, and $a$. As a point of comparison, it is useful to understand the static solution that is obtained by assuming the potential is minimized both inside and far outside the star, with the interpolating ``domain wall' profile found by minimizing the gradient energy of the field in the connecting region~\cite{Hook_Huang_2018,piinsky}. In the limit where the Compton wavelength of the axion is much shorter than the radius of the host NS, $m_a^{-1} \ll R_{\rm NS}$, the domain wall of the axion phase transition contributes to the total energy density and pressure in a shell that is negligibly thin. In this limit, the solution for the axion field is, to a good approximation, a step function~\cite{white-dwarf-2024,piinsky}. With this approximation of the axion profile, the static system of equations for the metric coefficients and nucleon number density can be solved using standard methods and the resulting effects on the NS calculated. The results presented in Ref.~\cite{piinsky} show that low-mass, high-radius NS are significantly affected by the presence of an LQCD axion, leading to a collapse of the outer crust of the star and a net suppression of the radius. We expect similar results in the case of an axion profile with a finite domain-wall width.

In the case where the axion domain wall has a finite spatial extent, it is in principle still possible to solve the static equations of motion using standard methods. Taking Eqs.~(\ref{eq:dFdr})-- (\ref{eq:axion-EOM-explicit}) and setting temporal derivatives to zero, one obtains the static system of four equations for the metric coefficients, nucleon number density, and axion profile. The initial conditions for all fields at a finite central radius $R_{NS} \gg r_c > 0$, where recall that $R_{NS}$ is the NS radius, are specified in terms of a central number density and a central axion field value. Given an initial condition on the radial evolution equations, one could then integrate from $r_c$ outwards to obtain the static solution. By evolving the fully dynamic system, we validate this approach by demonstrating that the dynamics of the neutron star damp away efficiently. We solve for the static solution by evolving the complete dynamic system numerically until an approximately static state is reached. We discuss this procedure in Sec.~\ref{sec:evolution-and-steady-state}.

\section{Numerical Methods}
\label{sec:numerical-methods}

In this section, we outline the numerical methods employed to evolve the combined axion and NS system. The core challenge of the system's evolution is the inherent numerical instabilities in the problem. As outlined below, this problem is resolved through a combination of a $4^{\rm th}$-order Runge-Kutta integration scheme and an artificial viscosity. While more advanced methods exist to resolve issues of numerical instability, they are beyond the scope of this work and would require dedicated analysis to implement.

\subsection{Explicit First Order Time Evolution}
As they stand, the equations of motion in Eqs.~(\ref{eq:dHdr})~-(\ref{eq:axion-EOM-explicit}) are difficult to analyze numerically because $dn_N/dt$ and $du/dt$ are dependent on time derivatives of the metric and matter fields.\footnote{In principle, one could evolve the system with implicit numerical methods, but we found these methods to be prohibitively slow.} It is therefore useful to introduce the scaled axion velocity $\Pi$, 
\begin{equation}
    \label{eq:field-redef}
    \dot{a} \equiv \frac{F}{H} \Pi\,,
\end{equation}
which simultaneously reduces the second-order equation of motion for the axion field to first order and produces expressions with explicitly defined time derivatives for the matter fields.
This parametrization obscures some of the intuitive dynamics in Eqs.~(\ref{eq:dFdr})--(\ref{eq:axion-EOM-explicit}), but it renders the system more tractable numerically. 

We evolve our differential system by adapting the numerical methods employed in Ref.~\cite{chop} to study critical gravitational collapse. In essence, the matter fields are stepped forward in time first using a 4$^{\rm th}$-order Runge-Kutta temporal integration scheme with 4$^{\rm th}$-order finite difference spatial derivatives. Subsequently, the metric coefficients are solved using the radial evolution equations with a spatial 4$^{\rm th}$-order Runge-Kutta scheme. We expand upon the techniques used for the matter and metric evolution below.

\subsection{Simulation Lattice}
We implement a radial grid with $N_r = 10^4$ points and a radial step size of $\Delta r = 10^{-2}~\mathrm{km}$. The grid ranges from the central radius $r_c = \Delta r / 2$ to $r_{\rm ext} = 100~\mathrm{km} + \Delta r / 2$. The smallest internal radius is fixed by the radial step size $\Delta r$ to $r_c = \Delta r / 2$. The maximum external radius is approximately ten times the characteristic radius of a NS, such that the metric has sufficient room to decay to approximately flat space, and so that simple external boundary conditions can be employed without loss of accuracy. The time step is fixed as $\Delta t = \Delta r$ to ensure that the Courant–Friedrichs–Lewy condition is met~\cite{LeVeque_2012} (i.e. the time step $\Delta t$ is smaller than the time for a wave to propagate across a spatial cell, which at minimum will be $\Delta r/c$). We further validate the choice of time step by simulating with smaller $\Delta t$ which produces the same results. The time step leads to a maximum resolvable frequency in the simulations, which will be relevant in analyzing the oscillations of the axion and nucleon fields. For a time step $\Delta t = \Delta r/c = 10^{-2}~\mathrm{km}/c \sim 3 \times 10^{-8}~\mathrm{s}$, the maximum frequency of the simulation is $f_{\rm max} \simeq 1 / \Delta t = 3.0 \times 10^7~\mathrm{Hz}$. A minimum resolvable frequency will be set by the run time of the simulation. We find that simulations are stable up to physical times $t_{\rm max} \simeq 300~\mu\mathrm{s}$, leading to a minimum resolvable frequency of $f_{\rm min} \simeq 3.3 \times 10^3~\mathrm{Hz}$.

\subsection{(Radial) Boundary Conditions}

Spherical coordinates require that functions be symmetric about the origin, and therefore, the radial derivative of all functions in the simulation must be zero at $r_c$. The boundary condition on all fields $f$ at $r_c$ in the simulation is
\begin{equation}
    \label{eq:boundary-conditions}
    \frac{\partial f}{\partial r}\bigg|_{r = r_c} = 0.
\end{equation}
The derivatives at the inner boundary are calculated using a $4^{\rm th}$-order forward finite difference and then applying the boundary condition in Eq.~(\ref{eq:boundary-conditions}). In practice, this means the central field values are defined by
\begin{equation}
    \label{eq:boundary-condition-numeric}
    f(r_c) = \frac{9 f(r_c + \Delta r) - f(r_c + 2 \Delta r)}{8}~_,
\end{equation}
which we obtain by taking the discrete first derivative, setting it to zero, and using the fact that the function is even to solve for $f(r_c)$.

We evolve the simulation to times long enough that waves can reflect off the outer boundary at $r_{\rm ext}$. To avoid any fictitious incoming waves, we employ a so-called ``sponge.'' In the final $N_s$ spatial cells of the simulation, we implement a damping term in the axion equation of motion that kills any outgoing oscillations, preventing them from reflecting from the outer boundary. The sponge must be turned on slow enough, so that there is no sponge reflection off the boundary. The sponge term added to the EOM for $\Pi$ is
\begin{equation}
    \label{eq:sponge}
    \mathcal{S} = -s f_s(r) \; \Pi \; \theta(r - r_s)
\end{equation}
for a sponge function
\begin{equation}
    \label{eq:sponge-function}
    f_s(r) = \frac{(r - r_s)^3 / \ell_s^3}{\sqrt{1 - (r-r_s)^6 / \ell_s^6}},
\end{equation}
where $s \gg 1$ is the damping ratio, $r_s$ is the radius at which the sponge term turns on, and $\ell_s$ is a characteristic length. In our simulations, we set $s = 1$, $r_s = 90~\text{km}$, and $\ell_s = 10~\text{km}$.

\subsection{(Temporal) Initial Conditions}
As we will demonstrate in Sec.~\ref{sec:evolution-and-steady-state}, the internal dynamics of all fields are rapidly and efficiently damped in physical times $\tau \sim \order{\mathrm{ms}}$. The youngest known NS is \textit{older} than at least $\order{10~\mathrm{yrs}}$~\cite{SN1987a,Cigan_2019}, so the macroscopic gravitational signatures of the axion field are all insensitive to the choice of initial conditions. In designing an initial state, we prioritize field profiles that lead to numerically-stable evolution. For example, an initial axion profile that everywhere starts far away from the value that minimizes the potential results in large changes in the axion field, leading to significant pressure gradients, shock formation in the matter fields, and numerical instability. It is unrealistic that we would observe any signature of this initial state in Nature due to the rapid decay of oscillations caused by the loss terms. Similarly, our simulations are not robust to the dynamics of star's formation, and we therefore choose solutions to the static, axion-free equations (i.e. the unmodified TOV equations) for the initial states of the nucleon fields.

To solve the axion-free system of equations for the initial state nucleon density profile, we integrate out from the center of the star. There is a coordinate singularity that prevents initial conditions for radial integration from being specified at the origin, so instead they are specified at a radius $r_c \gtrsim 0$ near the origin. The central nucleon density is fixed at $n_N(r_c) = n_N^c$ and the static equations can be expanded near $r_c$ to find the initial condition for $H$. To leading order, $H(t~=~0,r)~=~1~+~4~\pi~G_N~n_N^c~r_c^2~/~3~+~\order{r_c^4}$. With these initial values, Eqs.~(\ref{eq:dHdr}) and~(\ref{eq:dnNdt}) can be integrated out to $r_{\rm ext}$.

The choices of initial conditions for the axion $a$ and field velocity $\Pi$ are less prescriptive. The only firm requirement is that the initial state must satisfy the boundary conditions at the edges of the simulation box. To demonstrate that the final state is largely independent of the axion initial conditions, we simulate an ensemble of initial Gaussian profiles for two example stars, one with parameters that will source an axion field and one with parameters that will not. The final states of these simulations are shown in Appendix~\ref{app:numerics} and discussed in more detail there. Throughout the remainder of the paper, all simulations use an initial state axion profile defined by a step function with step at radius $r_{\rm crit}$ defined in terms of the critical nucleon number density by $n_N(r_{\rm crit}) = n_N^{\rm crit}$ and convolved with the window function $\mathcal{W}(r, w) = \Theta(r + w) - \Theta(r - w)$, namely
\begin{equation}
    \label{eq:axion-initial-state}
    a(t = 0, r) = \pi f_a \Theta\left( r_{\rm crit} - r \right) \otimes \mathcal{W}(r, 1~\mathrm{km}).
\end{equation}
We assume the field velocity vanishes everywhere at initialization, $\Pi(t = 0, r) = 0$.

\subsection{Stability}
The highly nonlinear nature of Eqs.~(\ref{eq:dFdr})--(\ref{eq:axion-EOM-explicit}) leads to instabilities in the numerical evolution of the system. We outline here the approaches necessary to mitigate and analyze the instabilities to ensure that the results of the simulations are robust.

The primary method for enhancing the stability of the simulations is an added artificial viscosity in the equations of motion. Absent this viscosity, we observe the creation and rapid growth of cell-to-cell oscillations in the numerical evolution of the system, particularly in the nucleon fields. We further observe the formation of shocks in the nucleon fluid arising from general initial conditions. To prevent the former issue and render the latter numerically viable, we add to each matter field equation of motion a damping term proportional to the second radial derivative of the field in accordance with the Lax-Friedrichs prescription~\cite{LeVeque_2012}. In order to prevent spurious flattening of the matter fields (i.e. to preserve the natural concavity of $f$), we apply the artificial viscosity to the high-frequency, short-wavelength modes by subtracting from the total field $f$, the value of $f$ convolved with a window function of width $2 \Delta r$:
\begin{equation}
    \label{eq:high-frequency-field}
    f_{\rm HF}(t, r) \equiv f(t,r) - f(t, r) \otimes \mathcal{W}(r, 2\Delta r)~.
\end{equation}
Additionally, we apply a cutoff to the artificial viscosity that limits the viscosity when the second spatial derivative of the field falls below a cutoff value $C_f$. Combining these conditions, for all matter fields $f$, we add the following damping
\begin{equation}
    \label{eq:LLF}
    \frac{\partial f}{\partial t} \rightarrow \frac{\partial f}{\partial t} + \frac{Q_f}{2} \Delta r \frac{\partial^2 f_{\rm HF}}{\partial r^2} \ \left(\frac{\mathrm{min}\left[ \left|\partial^2 f_{\rm HF}/\partial r^2 \right|, C_f\right]}{C_f} \right)^{\beta_f}~_.
\end{equation}
For each field $f$, the strength of the artificial viscosity is controlled by a dimensionless parameter $Q_f$. For spurious cell-to-cell oscillations of $f$, $\partial^2 f_{\rm HF} / \partial r^2$ will be large, and the artificial viscosity will damp away these oscillations. However, if $\partial^2 f_{\rm HF} / \partial r^2$ is below a characteristic value $C_f$, the viscosity will be continuously limited by a factor of $C_f^{-\beta_f} (\partial^2 f_{\rm HF} / \partial r^2)^{\beta_f}$. The factor of $\Delta r$ ensures that as the continuum limit is approached, the artificial viscosity vanishes. Importantly, the artificial viscosity only decreases as $\Delta r^2$ rendering the complete solution of the system only accurate to second order~\cite{LeVeque_2012}. In future work, we intend to implement more robust stabilizing methods that allow for fourth-order accurate solutions. We therefore still employ fourth-order Runge-Kutta integration in the time domain and fourth-order finite difference derivatives in the spatial variables but emphasize that at present the simulations are not accurate at fourth order.

The values of $Q_f$, $C_f$, and $\beta_f$ are chosen for each field to apply the minimum amount of artificial damping while still maintaining the stability of the simulation. For fields $f = \left\{n_N, u, a, \Pi \right\}$, we set $Q_f = \left\{ 0.05, 0.25, 0.25, 0.25\right\}$, $C_f = \big\{10^{-42}~\mathrm{GeV}^5, 10^{-49}~\mathrm{GeV}^{2}, 10^{-41}~\mathrm{GeV}^{3}, 10^{-57}~\mathrm{GeV}^{4}\big\}$, and $\beta_f = \left\{ 3, 1, 1, 1\right\}$. We verify that the results of the simulation are insensitive to the choices of viscosity parameters $Q_f$, $C_f$, and $\beta_f$, up to the point that numerical stability is achieved and spurious flattening of the fields is avoided.

To verify that the solutions are numerically independent of the integration method, we perform simulations with reduced temporal and radial time steps and find that the results do not differ significantly from the step sizes used in the results reported here. This comparison indicates that the system is numerically stable. Details can be found in Appendix~\ref{app:numerics}. While these methods for stabilizing the simulation are suitable for our analysis, more advanced and robust methods exist and will be employed in future work. Current state-of-the-art fluid simulations employ flux-limiting methods to enhance numerical stability and accuracy (see e.g. Refs.~\cite{clarisse2025,Castillo_2025} or Ref.~\cite{LeVeque_2012} for a technical overview).

\section{System Evolution and Steady State Solutions}\label{sec:evolution-and-steady-state} 
The analysis of the axion, metric, and NS dynamics spans a large parameter space: axion model parameters $\epsilon$ and $f_a$, nuclear EOS variation, and central nucleon density $n_N^c$. Throughout the study, we focus on a single nucleonic EOS, the SLy4 (zero-temperature, $\beta$-equilibrium) EOS. We analyze $100$ NSs with central nucleon number densities in the range $n_N^c~\in~\left[10^{-10},~10^{-9}\right]~n_N^{\rm sat}$ for nuclear saturation density $n_N^{\rm sat} = 1.2 \times 10^{-3}~\mathrm{GeV}^3$ to analyze the complete range of macroscopic NS observables. For each central density, we simulate axion fields with six sets of axion parameters listed in Figs.~\ref{fig:MR-curves} and \ref{fig:LC-curves}; this set spans the unexplored region of LQCD axion parameter space in Fig.~\ref{fig:LQCD_parameter_space} in which the axion profile cannot be modeled by a simple step function. In this section, we comment on the behavior of the dynamics observed throughout all simulations and analyze the steady-state behavior of the matter fields.\footnote{Animations of the axion and NS dynamics are available at $\texttt{github.com/wentmich/axions-in-NS}$.}
 
\subsection{Properties of the Dynamic Fields}
The initial state of the simulation does not minimize the axion potential, which immediately induces non-trivial dynamics in the matter and metric fields. Although the dynamics of the axion inside the star are highly nonlinear, insight can be gained through analysis of the dynamics of the axion field outside the star and the dynamics of the nucleon fields. 
\begin{figure}[t!]
    \centering
    \includegraphics[width=0.47\textwidth]{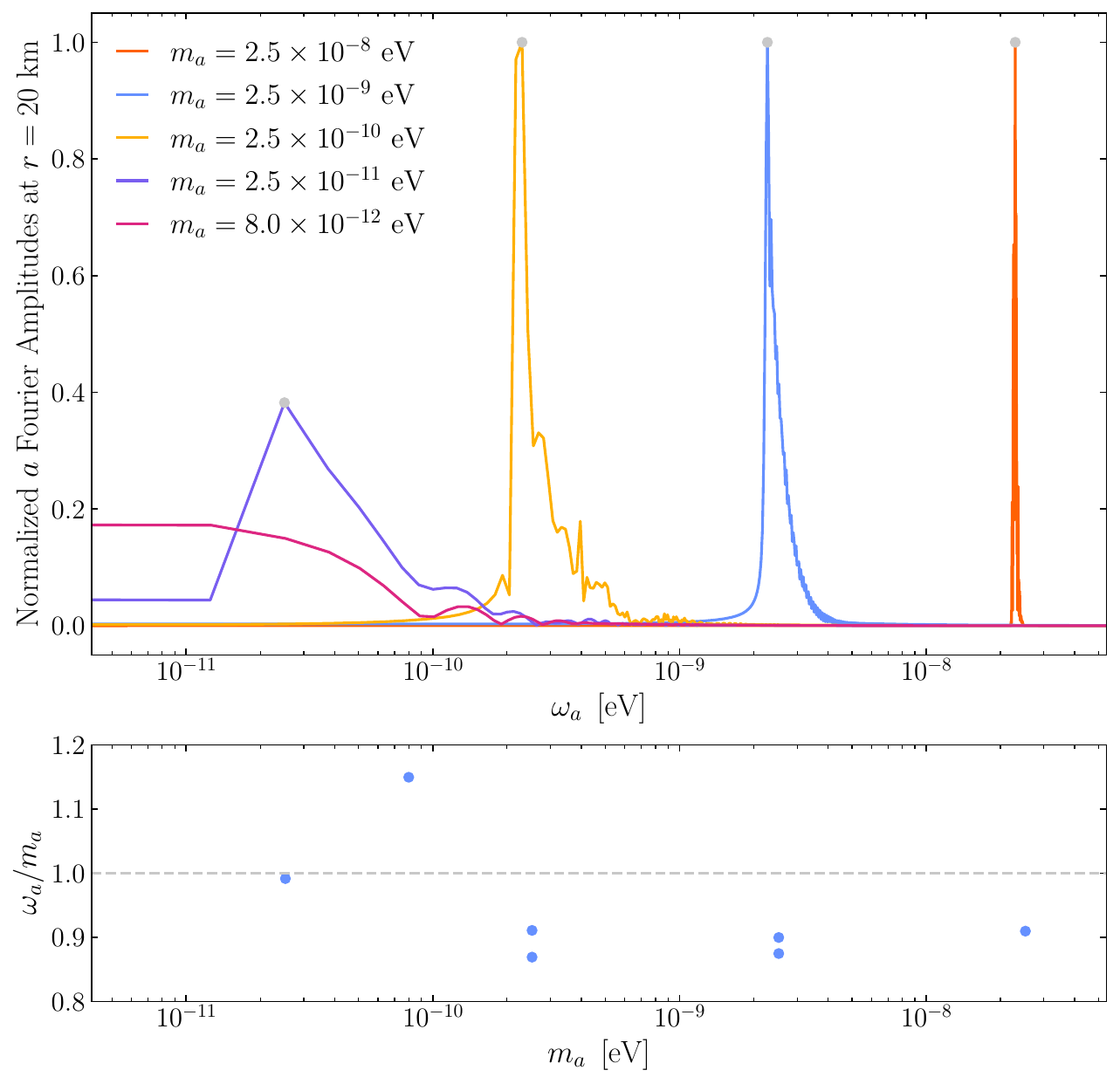} \\
    \caption{Normalized Fourier coefficients for the external oscillations of five representative axion fields at $r = 20~\mathrm{km}$, simulated with an initial state central baryon density $n_N^c = 5.0 \times 10^{-3}~\mathrm{GeV}^3$ (TOP). Peak frequencies for the seven axion parameter sets $(\epsilon, f_a)$ with initial central baryon density $n_N^c = 5.0 \times 10^{-3}~\mathrm{GeV}^3$ and resolvable frequencies given the simulation time step and run-time (BOTTOM). Ratio of the peak frequencies from the top panel to the axion mass, as a function of the latter. Observe that clearly the external oscillations occur at a frequency dictated by the vacuum axion mass, $\omega_a \simeq m_a = \sqrt{\epsilon} m_a^{\rm (QCD)}$.}
    \label{fig:axion-frequencies}
\end{figure}

At radii outside the region where $n_N > n_N^{\rm crit}$,  the axion field will fall off from $a = \pi f_a$ inside the star to an average value of $a = 0$. In this region, the equation of motion for the axion reduces to a massive Klein-Gordon equation in curved spacetime. We na\"ively expect any oscillations of the axion field outside the star to occur at frequencies $\omega_a \simeq m_a$. However, the complicated form of the axion EOM, including the damping/anti-damping term $\dot{a} F^{-2} \partial_t \log F / H$, indicates that nontrivial dynamics may occur. In Fig.~\ref{fig:axion-frequencies}, we plot the frequency of axion oscillations at a radius of $20~\mathrm{km}$, well outside the sourcing region of the axion, and we verify the na\"ive expectation, showing that for axion masses within the resolution of our simulations, the axion oscillates at frequency $\omega_a \simeq m_a$.
\begin{figure}[t!]
    \centering
    \includegraphics[width=0.47\textwidth]{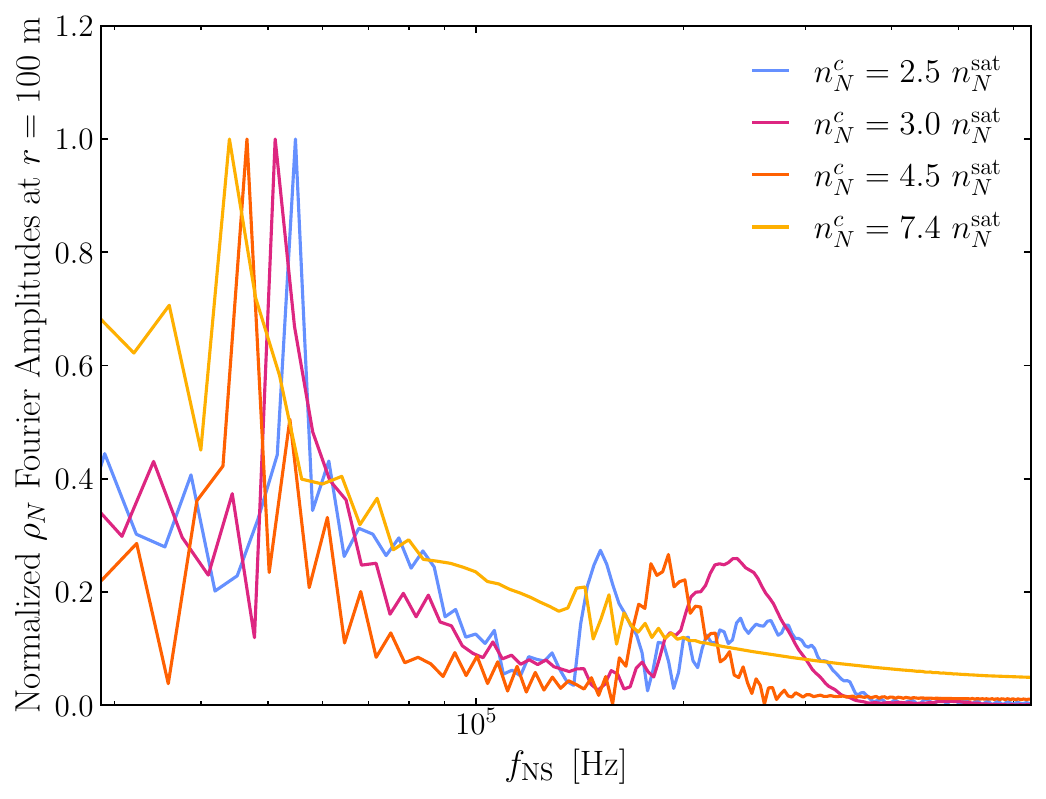} \\
    \caption{Normalized Fourier coefficients for the central oscillations of nucleon energy density $\rho_N$, simulated with four initial state central baryon densities $n_N^c \in \left\{ 2.5, 3.0, 4.5, 7.4 \right\}n_N^{\rm sat}$  with axion parameters $\epsilon = 0.1$ and $f_a = 10^{16}~\mathrm{GeV}$}
    \label{fig:nucleon-frequencies}
\end{figure}

As the nuclear matter responds to the axion oscillations, excitations in the normal modes of the NS will occur. Due to the assumed spherical symmetry of the system, these excitations will be entirely radial. The normal modes of the nucleon density depend strongly on the EOS, and a complete analysis of these modes is beyond the scope of  this work. However, it is interesting to observe the excitations of the fundamental modes of the star. From na\"ive dimensional arguments, the frequency of the fundamental mode should scale as $c_s / R_{\rm NS}$ where $c_s$ is the speed of sound in the star. For a NS with radius $R_{\rm NS}\sim 10~\mathrm{km}$ and central speed of sound $c_s\sim 0.7$, we expect mode excitations with frequencies $f_{\rm NS} \sim 20~\mathrm{kHz}$. Fig.~(\ref{fig:nucleon-frequencies}) shows the normalized Fourier coefficients of the nucleon energy density $\rho_N$ for eight central baryon densities $n_N^c \in \left\{ 2.5, 3.0, 4.5, 7.4 \right\} n_N^{\rm sat}$  with axion parameters $\epsilon = 0.1$ and $f_a = 10^{16}~\mathrm{GeV}$. It is evident that the nucleon density oscillates at a $\order{10~\mathrm{kHz}}$ frequencies, in line with the dimensional expectation. Note, however, that the na\"ive argument dictates that the frequency scale with the speed of sound when in fact, larger central densities (corresponding to higher $c_s$) lead to lower NS frequencies.

The EOMs in~(\ref{eq:dFdr})--(\ref{eq:axion-EOM-explicit}) also contain damping from the axion-metric coupling and from the added friction in Eq.~(\ref{eq:NS-normal-damping}). These sources of loss will cause the initially rapid oscillations of the simulation to decay to a steady state. The decay of the dynamics can be quantified by the kinetic contribution to the mass of the star. Given that the total energy density is
\begin{equation}
    \label{eq:rho-tot}
    \begin{aligned}
    \rho = &\frac{\rho_N + u^2 p_N(\rho_N)}{1 - u^2} + \frac{\dot{a}^2}{2F^2} + \frac{a'^2}{2 H^2} \\
    &+ \frac{\epsilon m_{\pi}^2 f_{\pi}^2 (1 - u^2) - \sigma_N n_N}{1 - u^2} \left( 1 - f(a) \right)~,
    \end{aligned}
\end{equation}
the kinetic contribution to the energy density can be defined using $\rho_{\rm tot}^{\rm kin} \equiv \rho_{\rm tot} - \rho_{\rm tot} (u = 0, \dot{a} = 0)$ via
\begin{equation}
    \label{eq:kinetic-energy-density}
    \begin{aligned}
    \rho_{\rm tot}^{\rm kin} = \frac{u^2}{1 - u^2} \left\{ \rho_N + p_N(\rho_N) - \sigma_N n_N \left[1 - f(a)\right]\right\} + \frac{\dot{a}^2}{2F^2}~.
    \end{aligned}
\end{equation}
The kinetic contribution to the mass is therefore
\begin{equation}
    \label{eq:kinetic-mass}
    M_{\rm NS}^{\rm kin} = \int_0^{R_{\rm NS}} 4 \pi r^2 \rho_{\rm tot}^{\rm kin} dr~.
\end{equation}
We restrict the integral to $r \leq R_{\rm NS}$ to avoid accounting for any axion and GW bursts propagating out of the star, as these will not affect the final state of the simulation. In the bottom-right panel of Fig.~(\ref{fig:steady-state-fields}), $M_{\rm NS}^{\rm kin}$ is plotted as a function of time for a characteristic simulation with $n_N^c = 2.7 \times n_{\rm sat}$, $\epsilon = 0.1$, and $f_a = 10^{16}~\mathrm{GeV}$. The decay rate of $M_{\rm NS}^{\rm kin}$ will depend on the form of the damping term for the nucleons in Eq.~(\ref{eq:NS-normal-damping}), but it is small enough that for reasonable choices of NS normal mode quality factor, the decay times will not be relevant on astrophysical time scales. 

In calculating gravitational observables of a NS, we will assume the static limit, where enough time has elapsed that $(\dot{a}, u) \rightarrow 0$. Practically, the dynamics of the simulation will never fully decay due to the artificial viscosity in Eq.~(\ref{eq:LLF}), which will eventually dominate the dynamics in the simulation. We therefore define the static limit of the simulation in terms of the fractional kinetic energy $\mathcal{F}_{\rm KE}$. We define the time $t_f$ when the simulation reaches the static limit by the time where the kinetic contribution to the mass decays by a factor of $100$ from its maximum value,
\begin{equation}
    \label{eq:steady-state}
    M_{\rm NS}^{\rm kin} = 0.01 \times M_{\rm NS}^{\rm kin, max}.
\end{equation}
This limit is chosen such that the energy density and pressure in the system are dominated by static contributions, indicating that the gravitational observables of the star have approximately stabilized. To ensure that the observables are stabilized, we further require that the discrete time derivative of the NS radius is zero to within the resolution of the simulation.

\subsection{Steady-State Matter Fields}
\begin{figure*}[t!]
    \centering
    \includegraphics[width=0.95\textwidth]{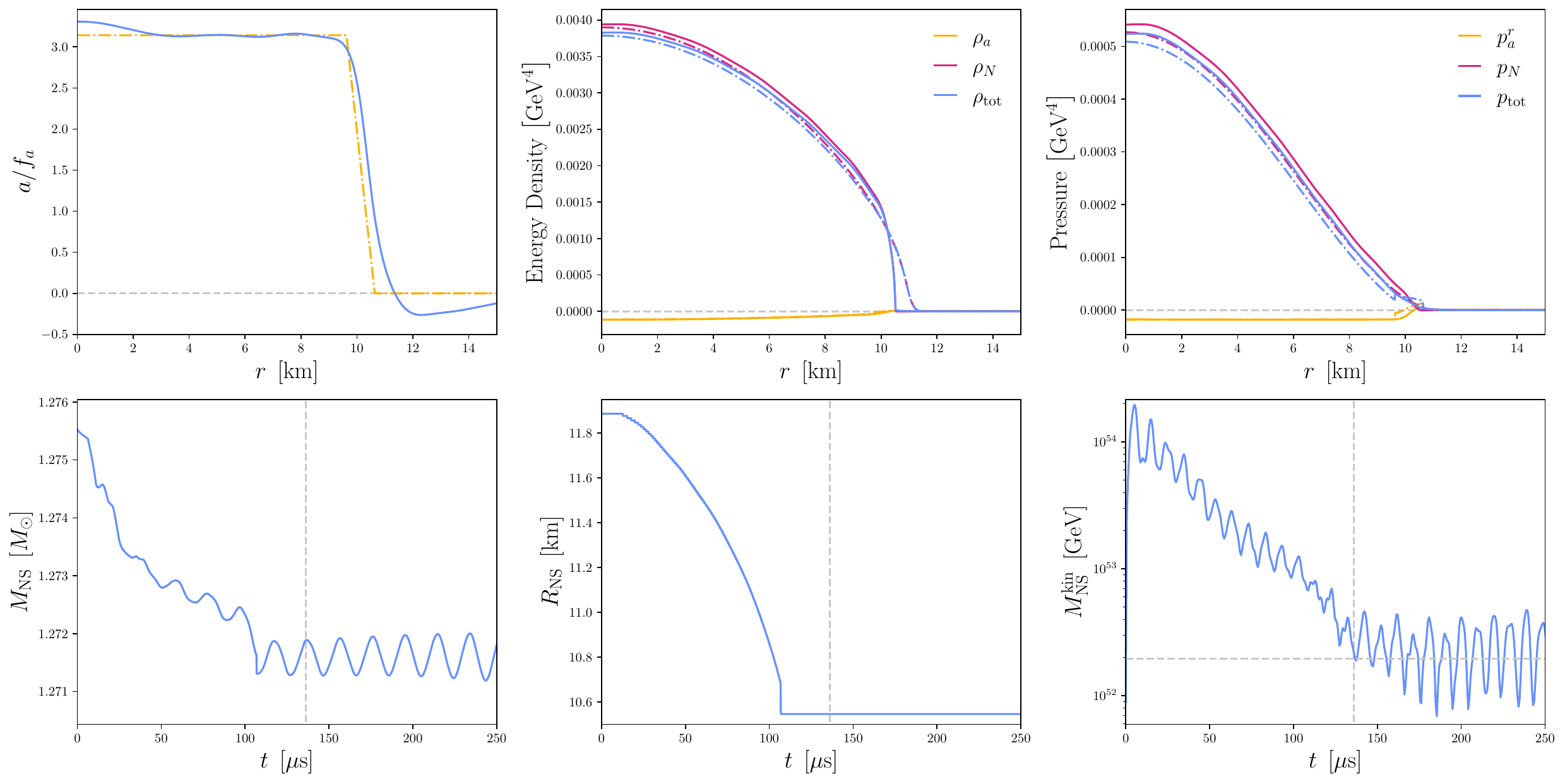} \\
    \caption{Initial state (dashed-dotted) and final state (solid) axion, energy density, and pressure fields for a representative simulation with $n_N^c = 2.7 n_{\rm sat}$, $\epsilon = 0.1$, and $f_a = 10^{16}~\mathrm{GeV}$ (TOP). Neutron star mass as a function of time (BOTTOM LEFT), neutrons star radius as a function of time (BOTTOM CENTER), and fractional kinetic energy as a function of time (BOTTOM RIGHT) for the same simulation. The vertical dashed line shows the time at which the final state observables are calculated, determined by the time at which the simulation reaches an approximately static state.}
    \label{fig:steady-state-fields}
\end{figure*}
In Fig.~\ref{fig:steady-state-fields}, we plot initial fields and the final, approximately static fields defined by Eq.~(\ref{eq:steady-state}) for the same representative simulation: $n_N^c = 2.7 \times n_{\rm sat}$, $\epsilon = 0.1$, and $f_a = 10^{16}~\mathrm{GeV}$. The quantities relevant for calculation of macroscopic observables are the axion field $a$ and the energy density and pressure. We find that for the initial state described in Sec.~\ref{sec:numerical-methods}, the axion field induces a negative pressure, which then has the effect of reducing the total pressure inside the star. When this happens, the star begins to shrink so that its internal density increases, thus increasing the pressure to balance the gravitational contraction. The overall net effect is a reduction in the stellar radius for a fixed baryonic mass. \
\begin{figure}[t!]
    \centering
    \includegraphics[width=0.47\textwidth]{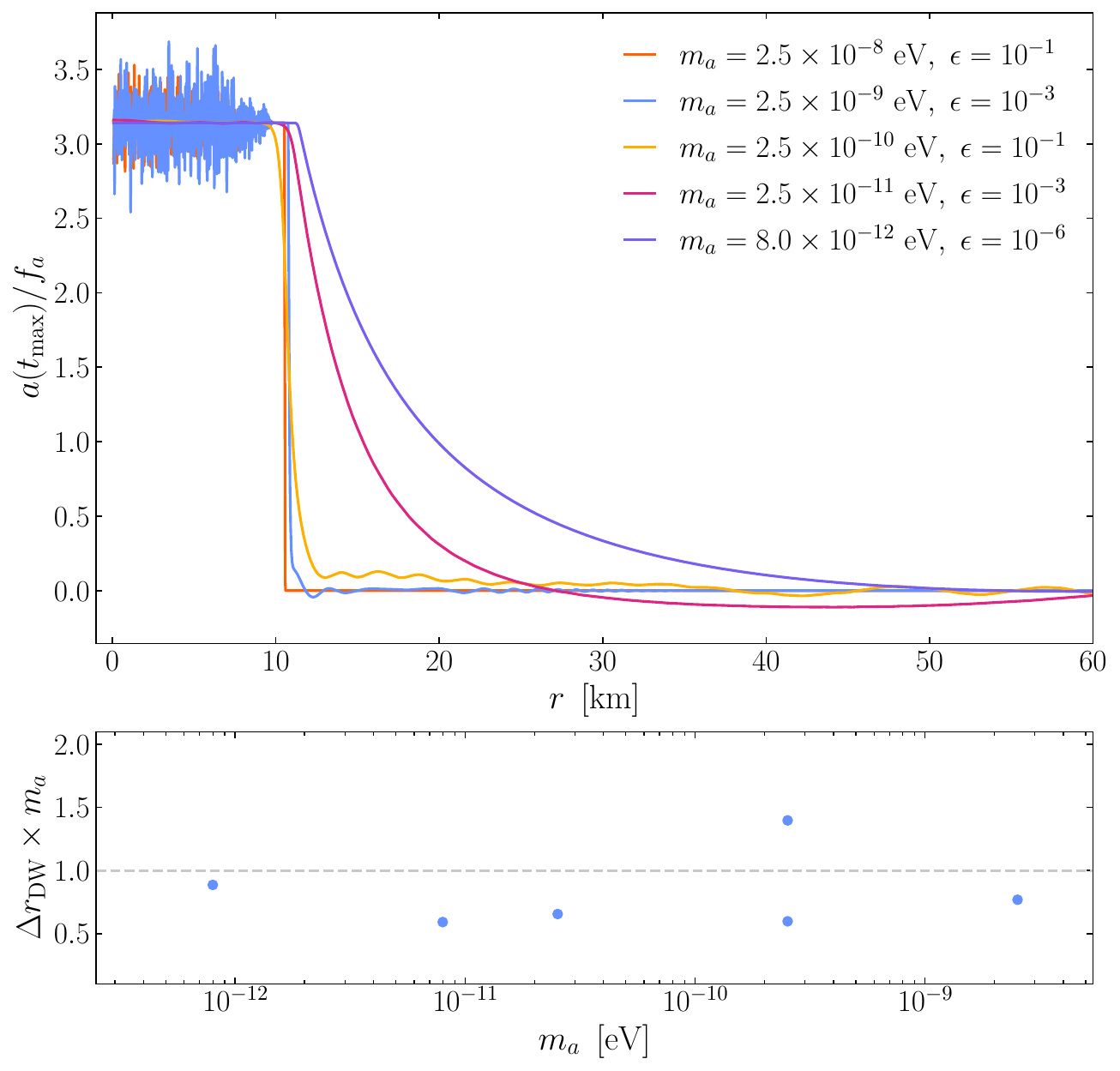} \\
    \caption{Final state axion field profiles for five axion masses simulated for an initial state central baryon density $n_N^c = 5.0 \times 10^{-3}~\mathrm{GeV}^3$ (TOP). Approximate fits for the domain wall width $\Delta r_{\rm DW}$ for all nine axion parameter sets $(\epsilon, f_a)$ for initial central baryon density $n_N^c = 5.0 \times 10^{-3}~\mathrm{GeV}^3$. The axion profile is approximated as $a / f_a \approx Q_{\rm eff} e^{-(r - r_0) / \Delta r_{\rm DW}} / r$, and the domain wall width is obtained by fitting the profile to this approximation (BOTTOM). The errors on the domain wall width fits are large, but one can see that the domain walls scale approximately as $m_a^{-1}$, independent of the values of $\epsilon$ and $f_a$ that produce $m_a$.}
    \label{fig:axion-DW}
\end{figure}

A sharp observer will notice that in the radial region very close to the surface of the star, the baryonic pressure becomes comparable in magnitude and then smaller than the axion pressure. In this regime, the axion cannot be treated as a perturbation to the baryonic configuration; non-linearities are important and must be treated carefully. We find that the final state has an axion pressure that decays more slowly than the baryonic pressure near the surface, forcing the total pressure to acquire a longer tail than in the axion-free case. This longer tail generates an ``axion atmosphere'' that extends beyond the baryonic radius (i.e.~the radius at which the baryonic pressure is zero, or very small relative to the central baryonic pressure.) 


The final state of the axion field matches expectations from previous studies~\cite{piinsky,white-dwarf-2024,Hook_Huang_2018}. It relaxes to $a = \pi f_a$ inside the star and falls to $a = 0$ outside, with an approximate domain wall located where $n_N$ drops below $n_N^{\rm crit}$. Due to the complex form of the potential near the surface of the NS, the functional form of the axion domain wall is correspondingly complex. However, the functional form of the domain wall can be approximated in the static limit by matching the interior solution that minimizes $V_{\rm eff}$ with a flat-space vacuum solution to the axion equation of motion on the exterior. With these simplifying assumptions, the approximate solution to the axion field profile is given in terms of a central axion value, a critical radius $r_{\rm crit}$ defining the location of the phase transition, and a  domain wall width $\Delta r_{\rm DW}$~\cite{Hook_Huang_2018}:
\begin{equation}
    \label{eq:axion-approximate}
    a(r) \approx a_c \Theta(r_{\rm crit} - r) + a_c r_{\rm crit} \frac{e^{-(r - r_{\rm crit}) / \Delta r_{\rm DW}}}{r} \Theta(r - r_{\rm crit}).
\end{equation}
For a flat spacetime, static axion field, $\Delta r^{\rm flat}_{\rm DW} = m_a^{-1}$. We should therefore expect the domain wall width to scale approximately as $m_a^{-1}$, an expectation we verify in Fig.~\ref{fig:axion-DW} for six sets of axion parameters simulated at a central density $n_N^c = 5.0 \times 10^{-3}~\mathrm{GeV}^3$. Note that for readability, in the top panel of Fig.~\ref{fig:axion-DW}, we show only five axion masses. The bottom panel highlights that the scaling is consistent across all simulations. The results of the simulation agree with expectation and validate the step-function approximation made in Refs.~\cite{piinsky,white-dwarf-2024} for larger $m_a$.

The representative simulation shown here depicts the typical response of the nucleons to the axion field based on the central nucleon density of the star and the axion parameters $\epsilon$ and $f_a$. The fact that the final state nucleon fields are influenced at an $\order{1}$ level by the axion field indicates that the macroscopic gravitational observables of the NS could be significantly altered in the presence of an axion. In the next section, we calculate these effects.

\section{Neutron Star Observables with Axions}\label{sec:observables}

The presence of a LQCD axion field inside and around a NS affects two key observables of the star: the mass-radius ($M$--$R$) relation and the tidal deformability-compactness ($\Lambda$--$C$) relation. These macroscopic NS observables have been studied extensively as tools to constrain the EOS for dense nuclear matter and to confront modified theories of gravity against Nature. Small changes to the EOS can lead to $\order{1}$ changes in the $M$--$R$ relation, making it a sensitive probe of the microphysics governing NS. Conversely, the tidal deformability-compactness relation is relatively insensitive to the EOS and can therefore be used as a model-agnostic probe of modified theories of gravity. The relations have been studied extensively in these respective contexts~\cite{Brandes_2025,Riley_2019,Riley_2021,Miller_2019,Miller_2021,Annala_2018,Bauswein_2017,Capano_2020,Dietrich_2020,Lattimer_2012,Romani_2022,Steiner_2013,yagi_2013_3,Ajith_2022,vylet,Yagi_2013,Yagi_2013_2,Nath_2023}, and the prospect of a significant increase in NS inspiral data from current and future GW detectors, along with improved results from NICER observations~\cite{Capote_2025,Riley_2021,Miller_2021}, promise improved $M$--$R$ and $\Lambda$--$C$ constraints.  

Here we use the results of our numerical simulations in Sec.~\ref{sec:numerical-methods} to analyze the effect of the axion on each of these observables. As in previous sections, we employ the zero-temperature, $\beta$-equilibrium SLy4 EOS for nuclear matter and use the $M$--$R$ and $\Lambda$--$C$ relations for pure (no-axion), SLy4 nuclear matter as a baseline comparison for the axion effects. Our goal here is not to directly set limits on LQCD axions based on existing NS observations, but rather, as a (hard but) first step, to illustrate how future NS analyses may be able to constrain the remaining LQCD axion parameter space.

\subsection{Mass-Radius Relation}\label{sec:MR}

The most straightforward effects of the axion on NS observables are the direct contribution of the (negative) axion energy density to the total NS mass (whose effect is to reduce to mass from the baseline), and the effect of the (negative) axion pressure on the NS radius (whose effect is to reduce to radius from the baseline). Given an EOS, a central density, an axion decay constant and mass, and an initial condition for the axion, the axion field can be calculated following the procedures set out in Secs.~\ref{sec:axion-metric-dynamics} and \ref{sec:numerical-methods}. For each simulation of a set of NS and axion parameters, the simulation is run until the steady state condition in Eq.~(\ref{eq:steady-state}) is met. Then, the enclosed mass function is calculated using Eq.~(\ref{eq:enclosed-mass}), while the baryonic radius of the star is defined by the radial coordinate at which the baryonic pressure vanishes. 

The observable effects of the axion on the mass of the NS is clear: the axion reduces the (ADM) mass of the star, defined by the integral of the total energy density (\ref{eq:rho-tot}) over the volume of the star
\begin{equation}
    \label{eq:total-mass}
    M_{\rm NS} = \int_0^{R_{\rm NS}} 4 \pi r^2 \rho dr~.
\end{equation}
This statement can be easily understood through an inspection of the axion energy density. Eq.~(\ref{eq:axion-energy-density}) and Fig.~\ref{fig:steady-state-fields} show that, in the approximately static limit, the dominant contribution to the axion energy density comes from the effective potential $V_{\rm eff}^{\rm LQCD}$. At radii $r < r_{\rm crit}$, the potential is strictly negative, thus suppressing the total energy density. Because the mass is the integral over the energy density, the total mass decreases. The only way to achieve a mass enhancement of the star is with an axion energy density dominated by the axion energy density terms proportional to derivatives of the axion field, $\dot{a}^2 / 2F^2$ and $a'^2 / 2H^2$. As discussed in Sec.~\ref{sec:evolution-and-steady-state}, the kinetic energy in the axion and nucleon fields inside the star will rapidly decay, and the radial derivative energy $a'^2 / 2H^2$ dominates only at the outer crust of the star for the choices of axion mass considered in this work. Since this occurs in a small shell in the crust of the star, the change of sign of the axion energy density cannot compensate the negative sign in the interior of the star, which thus leads to an overall mass reduction in general.

The observable effects of the axion on the radius of the NS depend on the method by which the radius is measured. Measurements of the NS radius can be made using observations of X-ray hot-spots on the NS surface~\cite{Riley_2019,Riley_2021,Miller_2019,Miller_2021}; in this case, assuming the X-rays are sourced by SM matter, the radius that is observed is the purely baryonic one, which is independent of the axion field profile outside the star. Alternatively, the NS radius may be measured indirectly through GW observations by first measuring the mass and tidal deformability of the star from the GW phase evolution, and subsequently inferring the radius. In this scheme, any energy density in the axion field will affect the inference of the radius. In either case, the radius $R_{\rm NS}$ is calculated as
\begin{equation}
    \label{eq:NS-radius}
    n_N(r=R_{\rm NS}) = n_N^{\rm cut}
\end{equation}
for a cutoff density $n_N^{\rm cut}$. We find that the two definitions of radius (i.e.~in terms of the baryonic pressure or the total pressure) are equivalent and insensitive to numerical artifacts, if we choose $n_N^{\rm cut} = 10^{-5} n_{N}^{\rm sat}$, with nuclear saturation density $n_{N}^{\rm sat} = 1.2 \times 10^{-3}~\mathrm{GeV}^3$. 

We find that the overall effect of the axion on the NS radius is an $\order{1}$ suppression throughout the axion parameter space we study. This effect is driven by the negative total system pressure induced by the axion near the surface of the star. The dynamics induced by the negative pressure result in the reduction of the NS radius, leading to a sharp drop in the density profile of the NS, as shown in Fig.~\ref{fig:steady-state-fields}. The effect is largest in systems where the region of negative pressure --- governed by the overlap region between where the axion is sourced and where the axion pressure dominates the total pressure --- is largest. The relative size of the effect is then governed by the pressure profile of the star, the density profile of the star, and the value of $\epsilon$ governing where the axion is sourced. For the SLy4 EOS employed in our analysis, we find that high values of the central nucleon density lead to density and pressure profiles with rapid decay near $R_{\rm NS}$. Conversely, stars with smaller central nucleon densities have density and pressure profiles that decay more slowly near $R_{\rm NS}$, leading to larger regions where the pressure can be dominated by the axion. We therefore expect the effect to be larger for low-mass, high-radius stars. The $\epsilon$ dependence of the radius suppression is complicated slightly by the fact that larger $\epsilon$ linearly increases the magnitude of the axion pressure, but also decreases the region in which the axion is sourced. In general, however, the radius suppression is expected to decrease with decreasing $\epsilon$ to satisfy the limit that as $\epsilon \rightarrow 0$, the zero-axion TOV solution is recovered.

\begin{figure}[t!]
    \centering
    \includegraphics[width=0.47\textwidth]{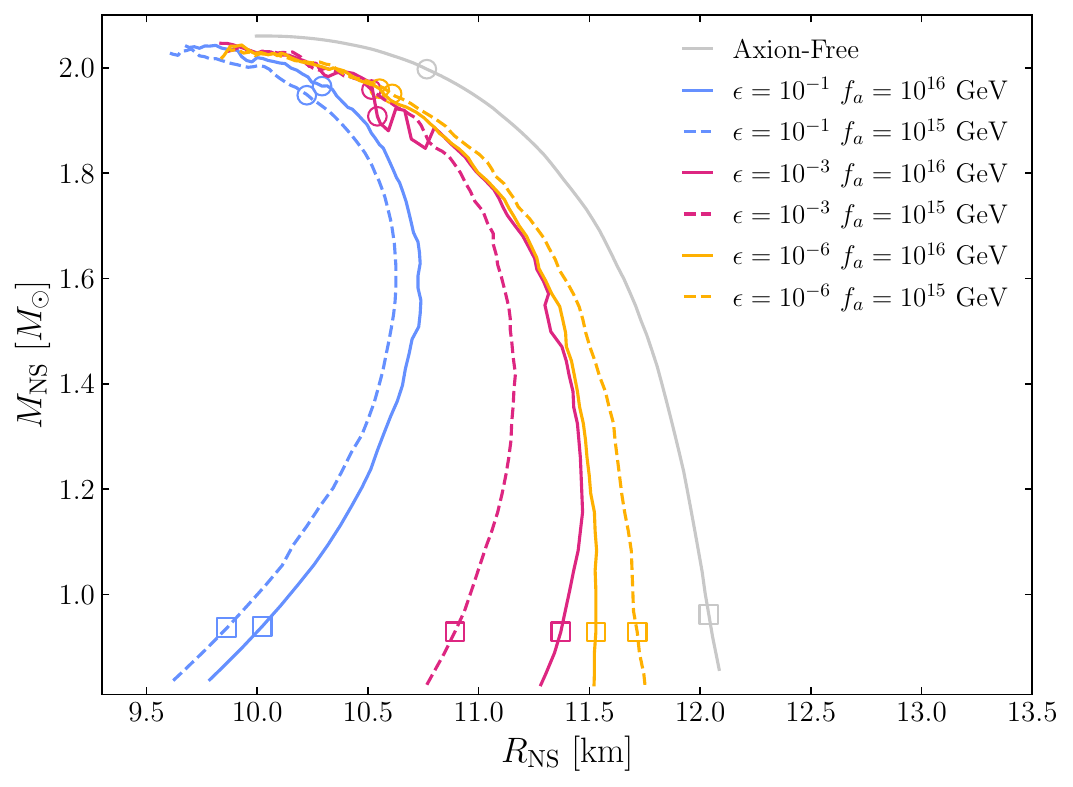} \\
    \caption{Mass-Radius curves for the purely-nucleonic model (gray) and with the addition of the LQCD axion for different choices of axion parameters $\epsilon$ and $f_a$. The addition of the axion results in a suppression of the radius that is largest for low mass/high radius stars and for large values of $\epsilon$. On each curve, two initial central nucleon number densities are shown: low density $n_N^c = 1.4 \times 10^{-3}~\mathrm{GeV}^3$ in boxed symbols, and high density $n_N^c = 7.5 \times 10^{-3}~\mathrm{GeV}^3$ in circled symbols. The effect of the axion on the $M$--$R$ curve suppresses the mass at the $\order{0.1}$ level due to the nucleon mass suppression and suppresses the radius at $\order{1}$.}
    \label{fig:MR-curves}
\end{figure}

\begin{figure}[t!]
    \centering
    \includegraphics[width=0.47\textwidth]{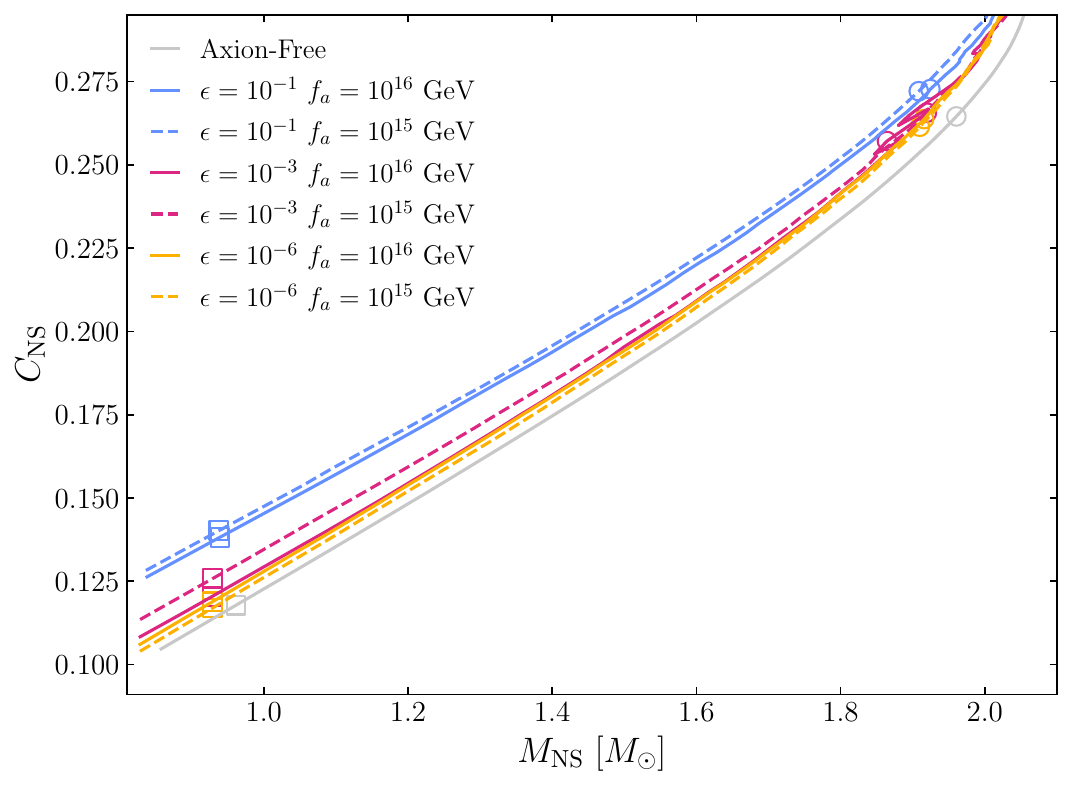} \\
    \caption{Compactness-Mass curves for the purely-nucleonic model gray) and with the addition of the LQCD axion for different choices of axion parameters $\epsilon$ and $f_a$. Relative to the axion-free star, the mass of the NS with an axion field will be marginally suppressed while the radius will be significantly reduced, in particular for large values of $\epsilon$. This effect results in stars with larger compactness for a given mass relative to the axion-free case.}
    \label{fig:CM-curves}
\end{figure}

The net effect of the axion on the NS $M$--$R$ curve is shown in Fig.~\ref{fig:MR-curves} with boxed and circled symbols representing two fixed values of the nucleon central density. Additionally, we show the effect of the axion on the $M$--$C$ relationship in Fig.~\ref{fig:CM-curves}. We see that the mass of the star is changed at the $\order{0.1}$ level, mainly due to the suppression of the nucleon mass shown in Eq.~(\ref{eq:mn-shift}). In accordance with expectation, the NS radii are suppressed in the presence of an axion field with larger suppression for lower mass NSs and larger values of $\epsilon$. Although this effect is significant, it is largely degenerate with the effects of an exotic nuclear EOS, for instance in strange quark stars. Indeed, strange quark stars have $M$--$R$ curves that resemble what we find here for low mass stars, but the radii of stars with larger masses are suppressed relative to strange quark stars~\cite{haensel_1986,Li_Yan_Ping_2021}.

\subsection{Tidal Deformability}
\label{sec:lambda-c}

\begin{figure}[t!]
    \centering
    \includegraphics[width=0.47\textwidth]{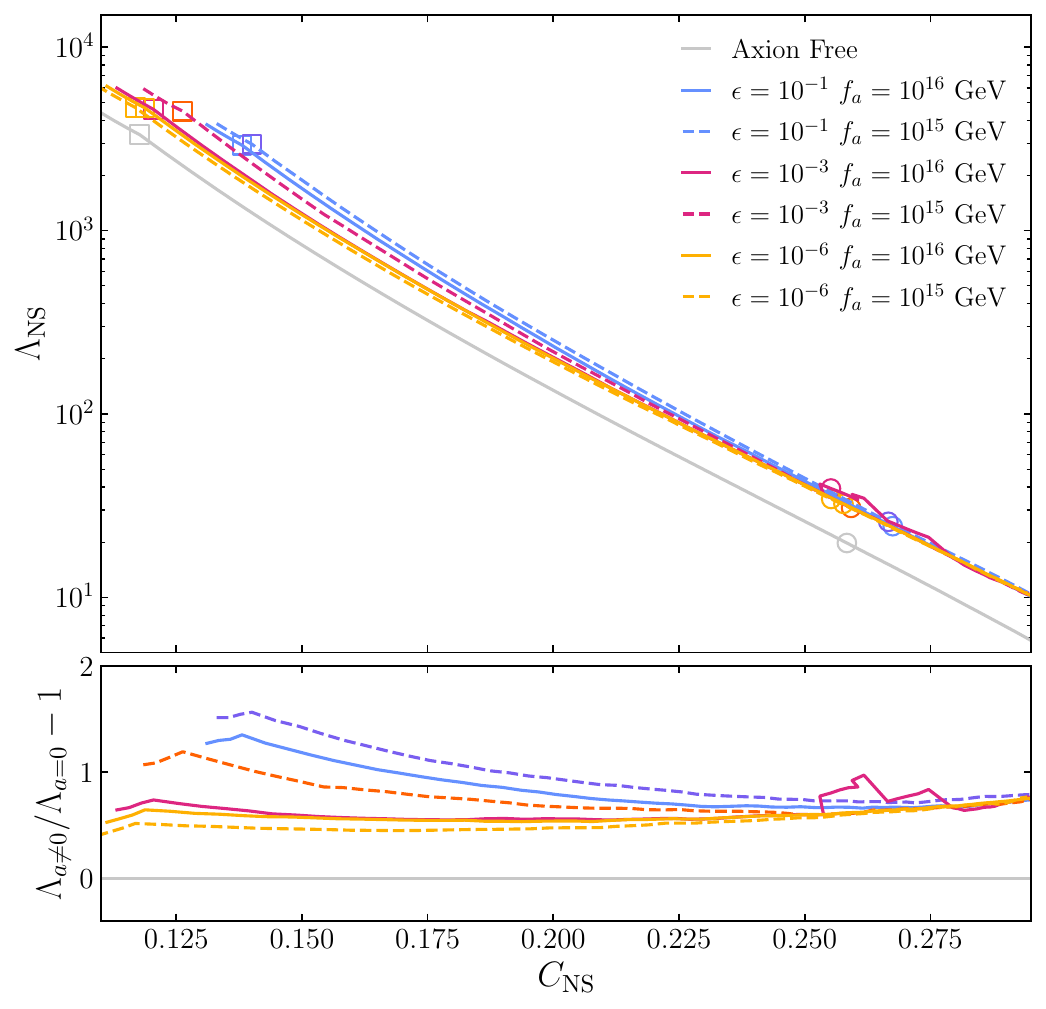} \\
    \caption{Tidal deformability-compactness curves for the purely nucleonic model (grey) and with the addition of the axion for difference choices of axion parameters $\epsilon$ and $f_a$ (TOP) and the differential effect of the axion, $\Delta \Lambda = \Lambda / \Lambda(a = 0) - 1$ (BOTTOM). The addition of the axion results in an increase of both the tidal deformability and the compactness of a star. On each curve, the effects for two central nucleon number densities are shown: low density $\rho_c = 1.9 \times 10^{-3}~\mathrm{GeV}^4$ in boxed points, and high density $\rho_c = 1.2 \times 10^{-2}~\mathrm{GeV}^4$ in circled points.}
    \label{fig:LC-curves}
\end{figure}

A LQCD axion also affects the quadrupolar tidal deformability $\Lambda$ of a NS, a measure of how the quadrupole moment of the star responds to quadrupolar tidal perturbations (for example, induced by a companion). This quantity is particularly relevant in the context of binary NS inspirals, where each NS tidally deforms the other, leading to a $\Lambda$-dependent phase shift of the GW waveform in the inspiral as energy dissipates from the system~\cite{flanagan-hindrer,Hinderer_2008,Chatziioannou_2020,Miller_Yunes_2021}, similar to how the tides of the Earth alter the Earth-moon orbit. In this section, we calculate the correction to $\Lambda$ in the presence of a LQCD axion and show that the axion alters the approximately EOS-invariant $\Lambda$--$C$ relation at an $\order{1}$ level, as shown in Fig.~\ref{fig:LC-curves}.

The standard formalism for calculating the tidal deformability assumes a quadrupolar perturbation $\mathcal{E}_0$ to a static and spherically symmetric star. In response to the perturbation, a quadrupole moment $\mathcal{Q}_0$ is induced in the star. The resultant spacetime is the sum of the two perturbations, which, in the intermediate (buffer) region between the star and the perturber, can be expanded as a bivariate series in powers of both $r$ and $1/r$. In spherical coordinates, the resultant metric coefficient $g_{tt}$ can be expressed in the local asymptotic rest frame of the star as~\cite{Thorne_1999,Hinderer_2008,flanagan-hindrer}
\begin{equation}
    \label{eq:metric-expansion}
    \frac{1 - g_{tt}}{2} = \frac{M_{\rm NS}}{r} - \frac{\mathcal{E}_0 r^2}{2} Y_{20} - \frac{3 \mathcal{Q}_0}{2 r^3} Y_{20} + \order{r^{-4}, r^3}~,
\end{equation}
where we have assumed that the perturbation does not break the polar symmetry of the system (because we are assuming non-spinning stars) to write the perturbation and response in terms of the $m=0$ (real) spherical harmonic $Y_{20}$. The quadrupolar tidal deformability $\lambda$ is defined by the leading-order relation between the induced moment and the perturbation: $\mathcal{Q}_0 = - \lambda \mathcal{E}_0$. The dimensionless tidal deformability $\Lambda$ is defined by normalizing by the total mass of the star, and is related to the dimensionless tidal Love number $k_2$ as
\begin{equation}
    \label{eq:Lambda-def}
    \Lambda = \frac{\lambda}{G_N^5 M_{\rm NS}^5} = \frac{2 R_{\rm NS}^5}{3 G_N^5 M_{\rm NS}^5} k_2 = \frac{2}{3 C_{\rm NS}^5} k_2~,
\end{equation}
where $C_{\rm NS} = G_N M_{\rm NS} / R_{\rm NS}$ is the compactness. Throughout, we will refer to $\Lambda$ simply as the tidal deformability.

To calculate the axion corrections to the tidal deformability, we will explicitly add a quadrupolar perturbation to the combined axion-NS spacetime and calculate the response. From Fig.~\ref{fig:steady-state-fields}, we see that, near the surface of the star, the axion and nucleon pressures are comparable, and therefore the effect of the axion cannot be calculated perturbatively. As such, we will calculate $\Lambda$ to leading order in tidal perturbations but to all orders in the axion field. As with the $M$--$R$ calculation, we determine the $\Lambda$--$C$ relation after the simulation has relaxed to the steady state, defined by Eq.~(\ref{eq:steady-state}). As in Eq.~(\ref{eq:metric-expansion}), we will expand functions $f_{nlm}(r, \theta, \phi)$ into real spherical harmonics such that $f_{nlm}(r, \theta, \phi) = f_{nl}(r) Y_{lm}(\theta, \phi)$ and select for the quadrupole $l = 2$ modes. Without loss of generality, we set $m = 0$ since the perturbations preserve the polar symmetry.

To begin, we add to the metric of a static and spherically symmetric star a spherical deformation to account for the tidal perturbations. The metric ansatz for a non-rotating but spherically deformed NS is
\begin{equation}
    \label{eq:metric-axion-perturbations}
    \begin{aligned}
    ds^2& = F^2 (1 + \kappa 2h_{2}(r) Y_{20}) dt^2  - H^2 (1 - \kappa 2h_{2}(r) Y_{20}) dr^2 \\
    & - r^2(1 - \kappa 2k_{2}(r)Y_{20}) d\theta^2 - r^2 \sin^2(\theta) (1 - \kappa 2k_{2}(r)Y_{20}) d\phi^2
    \end{aligned}
\end{equation}
where $\kappa \ll 1$ is a book-keeping parameter that represents the strength of the metric perturbation. The spherically-symmetric metric coefficients $F$ and $H$ are obtained directly from the LQCD simulations, while the functions $h_2$ and $k_2$ depend on the radius and must be solved for. 

The tidal perturbations will also affect the matter distribution, so an $\order{\kappa}$ quadrupolar perturbation to the stress-energy tensors will be induced:
\begin{equation}
    \label{eq:perturbed-Tb}
    \delta T^{(\mathrm{tot}) \mu}_{\quad \quad \nu} = \mathrm{diag}\big(\kappa \delta \rho, - \kappa \delta p, - \kappa \delta p, - \kappa \delta p \big).
\end{equation}
The perturbations to the stress-energy tensor depend on the perturbations to the density and the pressure, $\delta \rho$ and $\delta p$, which can be calculated from stress-energy conservation: $\nabla_{\mu} (T_{\rm tot}^{\mu\nu} + \delta T_{\rm tot}^{\mu\nu}) = 0$. In terms of the perturbations $h_2$ and $k_2$, the nucleon density and pressure, the axion density and pressure, and the total cumulative mass function $M_{\rm NS}$ defined by Eq.~(\ref{eq:total-mass}), these perturbations are given by
\begin{widetext}
\begin{gather}
    \label{eq:delta-rho}
    \delta\rho = \frac{(r - 2G_N M_{\rm NS}) \left[ 2 h_2 (p_a^{(\Omega)} - p_a^{(r)}) + r h_2 (\rho' + \rho_a') + 2 r k_2'( p_a^{(r)} - p_a^{(\Omega)} ) \right]}{G_N M_{\rm NS} + 4\pi r^3G_N p^{(r)}},\\
    \label{eq:delta-p}
    \delta p = - h_2 ( \rho + p^{(r)} ).
\end{gather}
Given these expressions, the $\order{\kappa}$ Einstein equations,
\begin{equation}
    \label{eq:delta-EEs}
    \delta G_{\mu\nu} = 8\pi G_N \delta T^{\rm (tot)}_{\mu \nu}~,
\end{equation}
yield a system of second-order differential equations for the tidal metric perturbations $h_2$ and $k_2$. The combination of the $(r,\theta)$ and $(r,r)$ components of Eq.~(\ref{eq:delta-EEs}) yield a solution for $k_2$ in terms of $h_2$,
\begin{equation}
    \begin{aligned}
    \label{eq:k2}
    k_2(r) = &\frac{1}{2} h_2' G_N \left(4 \pi  p^{(r)} r^3+M_{\rm NS}\right) + h_2 \frac{G_N^2 \left[16 \pi ^2 \left(p^{(r)}\right)^2 r^6+4 \pi  M_{\rm NS} \left(\rho +3 p^{(r)} \right) r^3-M_{\rm NS}^2\right] + r^2}{r \left(r-2 G_N M_{\rm NS}\right)}\\
    & \quad \quad - h_2 G_N r \frac{2 \pi  \left( \rho +p^{(r)} \right) r^3+M_{\rm NS}}{r \left(r-2 G_N M_{\rm NS}\right)}~.
    \end{aligned}
\end{equation}
The $(t,t)$ component yields a second-order ODE for the perturbation $h_2$,
\begin{equation}
    \begin{aligned}
    \label{eq:h2}
    h_2''(r) &= \Big\{4 \pi r^5 h_2  \Big[ 6 p_N + 4 p_a^{(r)}+p_a^{(\Omega)} \left(2-32 \pi  r^2 G_N p^{(r)}\right)  +r \left(64 \pi ^2 r^3 G_N^2 \left(p^{(r)}\right){}^3 - 4 \pi  r G_N \left[9 p_N + 5 \rho + p_a^{(r)} \right] p^{(r)} + \rho' \right)\Big] \\
    & + 2 r^2 h_2 M_{\rm NS} \Big[16 \pi ^2 G_N^2 p^{(r)} \left(8 p^{(\Omega)}+7 p^{(r)} + 5\rho \right) r^4 - 2 \pi  G_N \left(5 \rho + 16 p^{(\Omega)} + 5 p^{(r)} + 4 r \rho '\right) r^2+3\Big] + 4 h_2  G_N^2 M_{\rm NS}^3 \\
    &- 2 r h_2' \left(r-2 G_N M_{\rm NS}\right) \Big[4 \pi r^4 \left(2 \pi r^2 G_N p^{(r)} \big( p^{(r)} - \rho \big) + p^{(\Omega)}\right) - G_N M_{\rm NS}^2 + M_{\rm NS} \left(r-2 \pi  r^3 G_N \left\{\rho + 4 p^{(\Omega)} - 3 p^{(r)} \right\}\right)\Big] \\
    & + 4 h_2  G_N r M_{\rm NS}^2 \left[2 \pi  r^2 G_N \left(5\rho + 12 p^{(\Omega)} + 3 p^{(r)} + 2 r \rho' \right) - 3 \right]  \Big\} \times \Big\{r^2 \left(r-2 G_N M_{\rm NS}\right){}^2 \left(4 \pi  p^{(r)} r^3 + M_{\rm NS}\right)\Big\}^{-1}.
    \end{aligned}
\end{equation}
As a consistency check, sending $(\rho_a, p_a^{(r)}, p_a^{(\Omega)}) \rightarrow 0$ one recovers exactly existing results from the literature for the differential equation that $h_2$ must satisfy without an axion field~\cite{flanagan-hindrer,Miller_Yunes_2021, Chatziioannou_2020}. 

Given the definition of the tidal deformability in Eq.~(\ref{eq:Lambda-def}) in terms of the metric coefficient $g_{tt}$ (\ref{eq:metric-expansion}), a solution for the metric perturbation $h_2(r)$ must be obtained. With this in mind, the perturbed $g_{tt}$ component can be expanded in the intermediate $r$ regime to calculate $\Lambda$ by isolating the $r^2$ and $r^{-3}$ coefficients and determine the tidal perturbation moment $\mathcal{E}_0$ and response moment $\mathcal{Q}_0$, respectively. Equation~(\ref{eq:h2}) is analytically soluble on the exterior of the star, where the cumulative mass function is constant and the pressures and densities drop to zero. We find
\begin{equation}
    \label{eq:h2-exterior}
    \begin{aligned}
    h_2(r > R_{\rm NS}) = &\bigg\{2 G_N M_{\rm NS} \left(r-G_N M_{\rm NS}\right) \left[3 r^2 \left(4 c_1+c_2-2 c_2 \log \left(1-\frac{2 G_N M_{\rm NS}}{r}\right)\right)-2 c_2 G_N M_{\rm NS} \left(3 r+G_N M_{\rm NS}\right)\right] \\
    & \quad \quad\quad \quad\quad \quad + 3 \left[c_2 \log \left(1-\frac{2 G_N M_{\rm NS}}{r}\right)-2 c_1\right] r^4\bigg\} \times \bigg\{2 r G_N^2 M_{\rm NS}^2 \left(r-2 G_N M_{\rm NS}\right)\bigg\}^{-1}.
    \end{aligned}
\end{equation}
The quantities $c_1$ and $c_2$ are integration constants for the second-order ODE that $h_2$ must satisfy, which must be determined by solving the differential equations for $h_2$ on the interior of the star and matching this to the exterior solutions at the boundary, $r = R_{\rm NS}$. In practice, the equation for $h_2$ on the interior can be reparameterized into a first-order ODE equation via the change of variables $y(r) = ({R_{\rm NS}}/{h_2}) ({dh_2}/{dr})$. The equation for $y$ derived from Eq.~\eqref{eq:h2} is
\begin{equation}
    \label{eq:y-interior}
    \begin{aligned}
    y'(r) = &\bigg\{256 \pi ^3 G_N^2 \left(p^{(r)}\right)^3 r^4-16 \pi^2  G_N p^{(r)} \left[p_N^{(r)} (y+9)-(y-5) \rho_N +(y+1) p_a^{(r)}-(y-5) \rho_a \right] r^2 \\ 
    & + 4 \pi \left(-y^2+y+4\right) p_a^{(r)} - 8 \pi  p_a^{(\Omega)} \left(16 \pi  G_N p^{(r)} r^2+y-1\right) r^5 -4 \left(y^2-1\right) G_N^2 M_{\rm NS}^3  \\
    & + r^2 M_{\rm NS} \bigg[32 \pi ^2 G_N^2 p^{(r)} \left(p_N (y+15)-(y-5) \rho_N +8 p_a^{(\Omega)}+(y+7) p_a^{(r)}-(y-5) \rho_a\right) r^4 -y^2-y+6 \bigg] \\
    & + 4 \pi  G_N M_{\rm NS}  r^4\left[p_N \left(4 y^2+y-21\right)+(y-5) \rho_N +8 (y-2) p_a^{(\Omega)}+(y (4 y-7)-5) p_a^{(r)} + (y-5) \rho_a - 4 r \rho'\right]   \\
    & + 8\pi G_N^2 M_{\rm NS}^2  r^3 \left[-2 p_N y^2+ p_N y - \rho_N y + 15 p_N + 5 \rho_N -4 (y-3) p_a^{(\Omega)}+((5-2 y) y+3) p_a^{(r)} - (y-5) \rho_a+2 r \rho'\right] \\
    & + 4 \pi r \rho' - 4 \pi p_N \left(y^2+y-6\right) + 2 G_N M_{\rm NS}^2 r \left( 2 y^2+y-6 \right) \bigg\} \times \bigg\{r \left(r-2 G_N M_{\rm NS}\right){}^2 \left(4 \pi  p^{(r)} r^3+M_{\rm NS}\right)\bigg\}^{-1}
    \end{aligned}~_.
\end{equation}
and it can be solved numerically more easily in the interior of the star. Solutions for the integration constants $c_1$ and $c_2$ in Eq.~(\ref{eq:h2-exterior}) can then be obtained by matching $y(r)$ on the interior and exterior at the NS radius $R_{\rm NS}$ defined by Eq.~(\ref{eq:NS-radius}). We denote the matching value by $y^* = y(R_{\rm NS})$, in terms of which, we find
\begin{equation} 
    \label{eq:c2-solution-main}
    \begin{aligned}
    c_2 = &c_1 \bigg\{2 G_N M_{\rm NS} \left[4 (y^*+1) G_N^4 M_{\rm NS}^4+2 (3 y^*-2) G_N^3 R_{\text{NS}} M_{\rm NS}^3 + 2 (13-11 y^*) G_N^2 R_{\text{NS}}^2 M_{\rm NS}^2+3 (5 y^*-8) G_N R_{\text{NS}}^3 M_{\rm NS}\right] \\
    &\quad -6 G_N M_{\rm NS} R_{\text{NS}}^4 (y^*-2) -3 \log \left(1-\frac{2 G_N M_{\rm NS}}{R_{\text{NS}}}\right) R_{\text{NS}}^2 \left(R_{\text{NS}}-2 G_N M_{\rm NS}\right){}^2 \left[(y^*-2) R_{\text{NS}}-2 (y^*-1) G_N M_{\rm NS}\right] \bigg\}^{-1} \\
    & \quad \times \bigg\{6 R_{\text{NS}}^2 \left(R_{\text{NS}}-2 G_N M_{\rm NS}\right){}^2 \left(2 (y^*-1) G_N M_{\rm NS}-(y^*-2) R_{\text{NS}}\right)\bigg\}
    \end{aligned}
\end{equation}
The tidal deformability can then be calculated in terms of $y$ instead of $h_2$ by solving for $c_{1,2}$, using matching conditions on $y(r)$ at $R_{\rm NS}$, and one finds that the tidal deformability is given in terms of $y^*$.
\begin{equation}
    \label{eq:lambda}
    \begin{aligned}
    \Lambda_{\rm NS} = &\frac{16}{15} \bigg\{R_{\text{NS}}^2 \left[R_{\text{NS}}-2 G_N M_{\rm NS}\right]^2 \left[2 (y^*-1) G_N M_{\rm NS}-(y^*-2) R_{\text{NS}}\right] \bigg\} \\
    &\times \bigg\{2 G_N M_{\rm NS} \left[4 (y^*+1) G_N^4 M_{\rm NS}^4+2 (3 y^*-2) G_N^3 R_{\text{NS}} M_{\rm NS}^3+2 (13-11 y^*) G_N^2 R_{\text{NS}}^2 M_{\rm NS}^2+3 (5 y^*-8) G_N R_{\text{NS}}^3 M_{\rm NS}\right] \\
    & -6 (y^*-2) R_{\text{NS}}^4 G_N M_{\rm NS}  -3 \log \left(1-\frac{2 G_N M_{\rm NS}}{R_{\text{NS}}}\right) R_{\text{NS}}^2 \left[R_{\text{NS}}-2 G_N M_{\rm NS}\right]^2 \left[(y^*-2) R_{\text{NS}}-2 (y^*-1) G_N M_{\rm NS}\right]\bigg\}^{-1}.
    \end{aligned}
\end{equation}
\end{widetext}

A standard way to probe or constrain modified gravity is through the relation between the tidal deformability and the compactness, $C_{\rm NS}$, of a star. This is because the $\Lambda$--$C$ relation has been shown to be approximately universal for a large class of nuclear EOS~\cite{Yagi_2013,Yagi_2013_2,Nath_2023}. This means one can marginalize over our ignorance of the precise form of the nuclear EOS, and use the $\Lambda$--$C$ relation to test modified gravity~\cite{Gupta_2017,yagi_2013_3,Ajith_2022,vylet} and, in our case, to probe for the LQCD axion. Figure~\ref{fig:LC-curves} shows how the axion modifies this relation at $\order{1}$. Both the compactness of the star and the tidal deformability of the star are altered by the sourcing of an axion. The axion correction to the tidal deformability manifests as an increase in $\Lambda_{\rm NS}$ and a suppression in $C_{\rm NS}$, moving the $\Lambda$--$C$ curves up and to the right.  Physically, this makes sense: the axion reduces the mass of the star, making it easier to deform, and thus, inducing a larger quadrupole moment and a larger $\Lambda_{\rm NS}$. Moreover, the axion decreases the radius of the star, but not as much as it reduces the mass, and so the compactness increases, as depicted in Fig.~\ref{fig:CM-curves}. This effect remains consistent across stars of all central baryon densities, as depicted by the sets of boxed and circled symbols in Fig.~\ref{fig:LC-curves}. As was the case for the $M$--$R$ relation, the relative effect of the axion on the $\Lambda$--$C$ relation is larger for stars with smaller central nucleon number densities, a result of the fact that the change in $\Lambda$--$C$ is dominated by the compactness suppression.

Although the relative effect of the axion on the $\Lambda$--$C$ relation is of the same order as the $M$--$R$ effect, it is significantly more promising as a probe of the LQCD axion due to the approximate insensitivity of $\Lambda$--$C$ with changing nuclear EOS. Although current constraints on $\Lambda$--$C$ are weak~\cite{Silva_2021}, the promise of more and improved data in the coming years~\cite{Riley_2021,Miller_2021,Capote_2025} could lead to constraints on the untested LQCD parameter space.

\section{Conclusions and Outlook}
\label{sec:outlook}
In this paper, we have demonstrated that the macroscopic gravitational observables of NS are affected at $\order{1}$ by the presence of a LQCD axion, sourced from a finite nucleon density, an effect entirely independent of an initial macroscopic axion population. Our results indicate that, while no new constraints may immediately be placed on the LQCD axion, near-future measurements of the approximately EOS invariant $\Lambda$--$C$ relation may probe a previously untested region of LQCD $m_a$--$f_a$ parameters. In a theory-agnostic sense, this result is a nontrivial demonstration of the violation of the $\Lambda$--$C$ relation for NS when one considers additional non-baryonic fields. Therefore, our work is a promising indicator that the $\Lambda$--$C$ relation may be used in the future as a probe of various BSM effects. The specific $\Lambda$--$C$ signature of different BSM physics is certainly non-trivial and would require dedicated research.

In deriving these results, we have made several assumptions regarding the NS physics. While we expect these results to be qualitatively robust to changes in these assumptions, a more detailed analysis will be the subject of future work. In particular, we have studied a single nuclear EOS in this work, and while the $\Lambda-C$ relation is approximately EOS-invariant, the effects of the axion and the nuclear EOS on the $M-R$ relation are largely degenerate. The extent of that degeneracy, however, is marginal for low-mass, low-radius stars with $\epsilon \gtrsim 0.1$. Through a careful study of alternative nuclear EOS, we intend to understand better the overlap between these phenomena. Additionally, our treatment here, while fully nonlinear in the time-domain, still assumes a spherically symmetric spacetime. Introducing non-zero rotational and higher-order vibrational modes of the NS could lead to interesting phenomenology when coupled to an axion. We intend to further develop the numerical formalism employed in this work to study these effects in detail.

There are several other promising extensions of this work, specific to axion physics. The simulation framework we present here incorporates the complete nonlinear dynamics of the axion, nucleon fields, and metric coefficients along with a general nuclear EOS. This indicates that a study of the effects of a dynamic pion mass on realistic nuclear EOS, and subsequent effects on NS observables, is within reach, possibly enabling probes of LQCD axions with $\epsilon$ near unity. Additionally, we find that oscillations of the axion field outside the star are persistent, pointing to the possibility of electromagnetic signals from axion conversion in the magnetosphere; however, significant additional modeling is required to analyze both these possibilities. NSs thus remain an intriguing laboratory to study BSM physics, and our work extends the range of masses and couplings for which NSs can serve as a probe for axions.\newline

\noindent{\bf{Note Added}}: While this work was in preparation, Gómez‑Bañón, Pnigouras, \& Pons \cite{gomez} studied neutron stars with axion condensates by constructing static TOV+Klein–Gordon backgrounds and performing a linear perturbation analysis of radial ($\ell=0$) modes, identifying fluid‑ and axion‑dominated families and axion‑induced damping. Their analysis does not include tidal Love numbers, the $\Lambda$--$C$ relation, or non‑linear time‑domain dynamics. By contrast, our study evolves the coupled axion–NS system in 1+1 GR with a realistic EOS, explicitly includes anisotropic axion stresses, and computes the tidal deformability and its approximately EOS‑universal relation with compactness, finding order‑unity deviations and providing a closed‑form expression for $\Lambda$ [Eq.~(\ref{eq:lambda})] and $\Lambda$--$C$ curves [Figure~\ref{fig:LC-curves}]. The two approaches are, thus, complementary: theirs is seismology‑focused and linear, whereas ours targets GW‑facing tidal observables and the non‑linear backreaction that controls them.

\section{Acknowledgements}\label{sec:acknowledgements}
We thank Masha Baryakhtar, Carlos Conde, Will East, Junwu Huang, Luis Lehner, Cole Miller, D\'ebora Mroczek, Jaki Noronha-Hostler, Frans Pretorius, and Chloe Richards for helpful discussions. M.~W.~acknowledges support from the National Science Foundation (NSF) Graduate Research Fellowship Program, under Grant No. DGE 21-46756. Y.~K.~is supported, in part, by a Discovery Grant from the Natural Sciences and Engineering Research Council of Canada (NSERC). N.~Y.~acknowledges support from the Simons
Foundation through Award No.~896696, the Simons Foundation International through Award No.~SFI-MPS-BH-00012593-01, the NSF through Grants No.~PHY-2207650 and~PHY-25-12423, and NASA through Grant No. 80NSSC22K0806.

\bibliography{bib}{}

\newpage
\onecolumngrid

\appendix
\section{Lagrangian Derivation of the Nucleon Stress-Energy Tensor}\label{app:nucleon-SET}
In this appendix, we derive the perfect fluid stress energy density from the Lagrangian density presented in Eq.~\eqref{eq:LB}, following mostly the derivation in~\cite{Hawking_Ellis_1973}. Consider then a perfect fluid of nucleons of mass $m_N$ with energy density $\rho$, number density $n_N$, and pressure defined by the equation of state $p = p(\rho)$. The energy density can be decomposed into a sum of the rest mass energy and the number-density-dependent fractional internal energy $\xi(n_N)$ (per rest mass energy):
\begin{equation}
    \label{eq:rho-PF}
    \rho = m_N n_N \left[ 1 + \xi(n_N) \right]~.
\end{equation}
The pressure, number density, and fractional internal energy are related through the definition of pressure in terms of the energy density: 
\begin{equation}
p = n_N^2 \frac{d(\rho/n_N)}{dn_N} = m_N n_N^2 \frac{d\xi}{dn_N}\,,
\end{equation}
as explained in Refs.~\cite{compose,Hawking_Ellis_1973}.

The stress-energy tensor (SET) for the perfect fluid is derived from Eq.~(\ref{eq:EH-SET}). Applying Eq.~(\ref{eq:EH-SET}) to Eqs.~(\ref{eq:LB}) and (\ref{eq:rho-PF}), the following expression is obtained:
\begin{align}
    \label{eq:nucleon-SET-1}
    T^{(N)}_{\mu\nu} &= g_{\mu\nu} m_N n_N \left[ 1 + \xi(n_N) \right] - 2 m_N \frac{\delta}{\delta g^{\mu\nu}} \left\{ n_N \left[ 1 + \xi(n_N) \right] \right\} \\
    \label{eq:nucleon-SET-2}
    &= g_{\mu\nu} m_N n_N \left[ 1 + \xi(n_N) \right] - 2 \frac{\delta n_N}{\delta g^{\mu\nu}} \left\{ m_N \left[ 1 + \xi(n_N) \right] + m_N n_N \frac{d \xi}{dn_N} \right\}.
\end{align}
The first term of the above equation arises naturally, but the second term requires more thought. Particle number conservation is required to calculate the variation of the number density with respect to the metric, because the number density depends on the volume of the system, which, in turn, depends on the metric. We will show explicitly below how to calculate this variation.

The variation of the number density with the metric $\delta n_N / \delta g^{\mu\nu}$ is derived from particle number conservation, as given by the continuity equation $\nabla_{\mu} (n_N u^{\mu}) = 0$. Integrating the continuity equation over space and applying Stoke's Theorem, we can write the condition of particle number conservation in terms of an integral of the number current $n_N u^{\mu}$ over a hypersurface,
\begin{equation}
    \label{eq:integral-continuity-equation}
    \int \sqrt{-g} \nabla_{\mu}\left(n_N u^{\mu}\right) d^4x = \int n_N u^{\mu} \epsilon_{\mu\nu\sigma\lambda} e^{\nu}_1 e^{\sigma}_2 e^{\lambda}_3 d^3y = N_{\rm tot},
\end{equation}
where $\epsilon_{\mu\nu\sigma\lambda} = \sqrt{-g} \left[\mu\nu\sigma\lambda\right]$ is the Levi-Civita tensor, $e^{\alpha}_i = \partial x^{\alpha} / \partial y^i$, $y^i$ are the coordinates of the hypersurface, and $N_{\rm tot}$ is the total number of particles~\cite{Carroll_2024}. We require that the total particle number remain constant as the metric varies, and therefore
\begin{equation}
    \label{eq:current-epsilon-variation}
    \frac{\delta}{\delta g^{\alpha\beta}} \left(n_N u^{\mu} \epsilon_{\mu\nu\sigma\lambda} \right) = \frac{\delta n_N}{\delta g^{\alpha\beta}} u^{\mu} \epsilon_{\mu\nu\sigma\lambda} + \frac{\delta u^{\mu}}{\delta g^{\alpha\beta}} n_N \epsilon_{\mu\nu\sigma\lambda} + \frac{\delta \epsilon_{\mu\nu\sigma\lambda}}{\delta g^{\alpha\beta}} n_N u^{\mu} = 0~.
\end{equation}
After contracting Eq.~(\ref{eq:current-epsilon-variation}) with $u_{\mu}$ and applying the relations $\delta \epsilon_{\mu\nu\sigma\lambda}/\delta g^{\alpha\beta} = -(1/2)g_{\alpha \beta}\epsilon_{\mu\nu\sigma\lambda}$ and $u^{\mu} (\delta u^{\mu}/\delta g^{\alpha\beta})~=~(1/2)u_{\alpha}u_{\beta}$ (coming from the variation of the four-velocity normalization condition), we obtain
\begin{equation}
    \label{eq:number-density-variation}
    \frac{\delta n_N}{\delta g^{\mu\nu}} = - \frac{1}{2} n_N \left( u_{\mu} u_{\nu} - g_{\mu\nu} \right).
\end{equation}

With this result, Eq.~(\ref{eq:nucleon-SET-2}) becomes
\begin{equation}
    \label{eq:nucleon-SET-1}
    T^{(N)}_{\mu\nu} = g_{\mu\nu} m_N n_N \left[ 1 + \xi(n_N) \right] + \left( u_{\mu} u_{\nu} - g_{\mu\nu} \right) \left\{ m_N n_N \left[ 1 + \xi(n_N) \right] + m_N n_N^2 \frac{d \xi}{dn_N} \right\}~.
\end{equation}
After applying the definitions of pressure and energy density, the standard form of the perfect fluid SET is obtained, namely
\begin{equation}
    \label{eq:nucleon-SET-1}
    T^{(N)}_{\mu\nu} = u_{\mu} u_{\nu} (\rho + p) - g_{\mu\nu} p~.
\end{equation}

\section{Numerical Methods}\label{app:numerics}

\subsection{Time Stepping Procedure}
Following the method in Ref.~\cite{chop}, adapted slightly to fit our simulation, the procedure for time evolution of the simulation involves separately stepping the matter and metric fields using explicit time integration and solutions of the radial constraint equations respectively. The stepping of the explicit variables follows a $4^{\rm th}$-order Runge-Kutta scheme in time with $4^{\rm th}$-order finite difference radial derivatives. The constraint equations for $F$ and $H$ are solved using a $4^{\rm th}$-order Runge-Kutta scheme in radius.

To step the explicit variables forward in time, we define the initial state vector $\Vec{q} \equiv (\Vec{n_N}, \Vec{u^r}, \Vec{a}, \Vec{\Pi})$, where $\Vec{n_N}$, $\Vec{u}$, $\Vec{a}$, and $\Vec{\Pi}$ are $N_r$ dimensional vectors representing the functional values on the radial grid. The solution for the $(i + 1)^{\rm th}$ time step $\Vec{q}^{(i + 1)}$ is given in terms of $\Vec{q}^{(i)}$ by
\begin{equation}
    \label{eq:RK4-solutions}
    \Vec{q}^{(i + 1)} = \Vec{q}^{(i)} + \frac{1}{6} \Vec{k}^{(i)}_1 dt + \frac{1}{3} \Vec{k}^{(i)}_2 dt + \frac{1}{3} \Vec{k}^{(i)}_3  dt + \frac{1}{6} \Vec{k}^{(i)}_4 dt~.
\end{equation}
The RK4 stepping functions $\Vec{k}_j$ are defined in terms of the the explicit time derivatives $\partial \Vec{q} / \partial t$ in Eqs.~(\ref{eq:dnNdt},\ref{eq:dudt},\ref{eq:axion-EOM-explicit},\ref{eq:field-redef}):
\begin{equation}
    \label{eq:RK4-kj}
    \begin{aligned}
        \Vec{k}^{(i)}_1 &= \frac{\partial \Vec{q}}{\partial t}(\Vec{F}^{(i)}, \Vec{H}^{(i)}, \Vec{u}^{(i)}), \\
        \Vec{k}^{(i)}_2 &= \frac{\partial \Vec{q}}{\partial t}(\Vec{F}^{(i)}, \Vec{H}^{(i)}, \Vec{q}^{(i)} + \Vec{k}^{(i)}_1 dt/2), \\
        \Vec{k}^{(i)}_3 &= \frac{\partial \Vec{q}}{\partial t}(\Vec{F}^{(i)}, \Vec{H}^{(i)}, \Vec{q}^{(i)} + \Vec{k}^{(i)}_2 dt/2), \\
        \Vec{k}^{(i)}_4 &= \frac{\partial \Vec{q}}{\partial t}(\Vec{F}^{(i)}, \Vec{H}^{(i)}, \Vec{q}^{(i)} + \Vec{k}^{(i)}_3 dt).
    \end{aligned}
\end{equation}
The time derivatives of the matter fields depend on the matter fields themselves and the metric coefficients. When calculating the $(i+1)^{\rm th}$ step of the matter fields, $F$ and $H$ are fixed at $\Vec{F}^{(i)}$ and $\Vec{H}^{(i)}$, while the explicit variables are shifted for each stepping function $\Vec{k}^{(i)}_j$. 

After solving for the explicit variables at the $(i + 1)^{\rm th}$ time step, the metric coefficients are stepped forward in time, solving the radial equations in Eqs.~(\ref{eq:dFdr})--(\ref{eq:dHdr}) for $\Vec{F}^{(i+1)}$ and $\Vec{H}^{(i+1)}$. Equation~(\ref{eq:dHdr}) is independent of $F$, so it is integrated first. 
Using the initial condition 
\begin{equation}
    \label{eq:Hc-condition-with-axion}
    \begin{aligned}
        H(t, r_c) = 1 + \frac{2\pi G_N}{3} \Bigg( & \frac{2 \left[ u(t, r_c) p(\rho(n_N(t, r_c))) + \rho(n_N(t, r_c)) \right]}{1 - u^2(t, r_c)} + \Pi^2(t, r_c) \\
        & \ \ - \frac{2\left[1 - f(a(t,r_c))\right]\left[ \sigma_N n_N(t,r_c) - \epsilon m_{\pi}^2 f_{\pi}^2(1 - u^2(t,r_c)) \right]}{1 - u^2(t,r_c)} \Bigg).
    \end{aligned}
\end{equation}
Eq.~(\ref{eq:dHdr}) is integrated from $r_c$ to $r_{\rm ext}$ using a $4^{\rm th}$-order Runge-Kutta scheme. 

The initial condition on $F$ can be determined by integrating Eq.~(\ref{eq:dFdr}) outside the star. In the limit that $r_{\rm ext} \gg R_{\rm NS}$, $[\bar{\rho}, \rho_a, p(\bar{\rho})] \rightarrow 0$, and the enclosed mass approaches a constant. Integrating Eq.~(\ref{eq:dFdr}) in this limit, we obtain
\begin{equation}
    \label{eq:F-ic}
    F^2(r_{\rm ext}) = 1 - \frac{2 M(r_{\rm ext})}{r_{\rm ext}}
\end{equation}
where $M(r)$ is the enclosed mass at radius $r$ defined by
\begin{equation}
    \label{eq:enclosed-mass}
    M(r) = \int_0^r dr' 4 \pi r'^2 (\rho_a + \rho_N)~.
\end{equation}
In terms of the matter fields, the enclosed energy density is independent of $F$, so we can compute consistently the enclosed mass $M(r_{\rm ext})$ at $r_{\rm ext}$ given the solutions for $\Vec{q}^{(i+1)}$. Equation~(\ref{eq:dFdr}) is then integrated from $r_{\rm ext}$ to $r_c$ to obtain a complete solution for $F^{(i + 1)}$.

\subsection{Friction Dependence}
\begin{figure}[t!]
    \centering
    \includegraphics[width=1.0\textwidth]{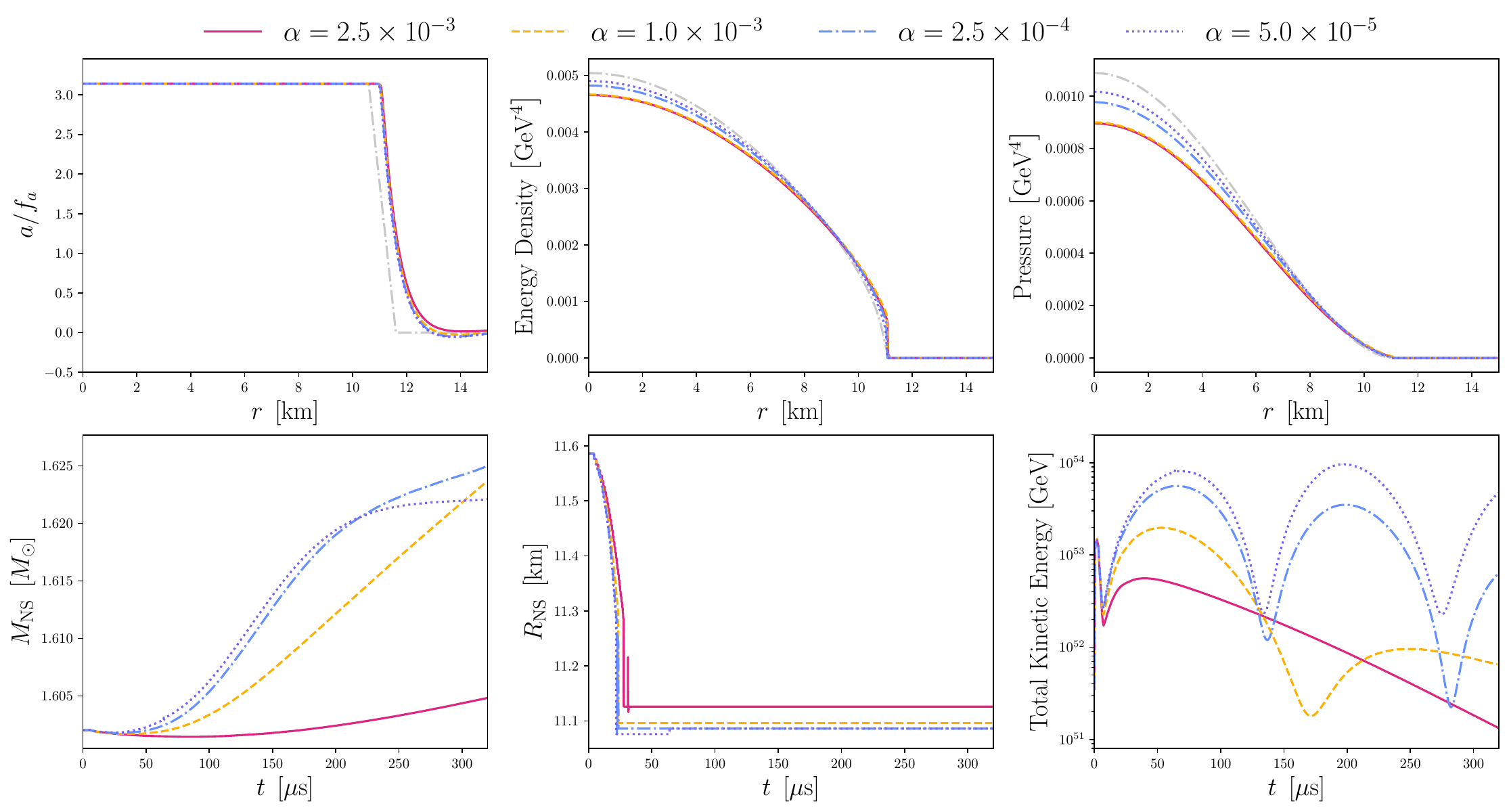} \\
    \caption{Test of the sensitivity of final state to friction coefficients $\alpha$ with axion parameters $\epsilon = 10^{-3}$ and $f_a = 10^{15}~\mathrm{GeV}$. Initial state (dot-dashed, grey) axion field and final state fields for four values of $\alpha$ spanning two orders of magnitude (TOP LEFT). Initial and final state total energy density (TOP CENTER) and total pressure (TOP RIGHT). Neutron star mass (BOTTOM LEFT), radius (BOTTOM CENTER), and total kinetic contribution to the NS mass (\ref{eq:kinetic-energy-density}) (BOTTOM RIGHT) as a function of physical time.}
    \label{fig:friction-test}
\end{figure}

Beyond the friction induced by the coupling of the axion to the metric in the axion EOM, it is expected that the additional damping will be present generically in the nucleon sector. This damping will result in the ringing and subsequent decay of the quasi-normal modes of the neutron star. While rigorous modeling of the NS normal mode damping is beyond the scope of this work, in this subsection, we verify that the observables analyzed throughout this study are approximately independent of the choice of damping rate. We model the nucleon sector damping through a frictional term in the equation of motion for $u$, defined via
\begin{equation}
    \label{eq:4-vel-app}
    u^{\mu} = \frac{1}{\sqrt{1 - u^2}} \left( \frac{1}{F}, \frac{u}{H}, 0, 0 \right)~_.
\end{equation}
The damping is modeled linearly as
\begin{equation}
    \label{eq:NS-normal-damping-app}
    \frac{\partial u}{ \partial t} \rightarrow \frac{\partial u}{\partial t} - \alpha \frac{u}{\Delta r}~_,
\end{equation}
where $\alpha$ is a dimensionless free parameter and $\alpha / \Delta r$ is the approximate decay rate of $u$. In Fig.~\ref{fig:friction-test}, we show the initial and final state axion field, total energy density, and total pressure. While the continued dynamics of the axion, density, and pressure fields lead to apparent differences in the fields for lower friction coefficients $\alpha$, the macroscopic observables are consistent across two orders of magnitude in $\alpha$. We see that for $\alpha \lesssim 10^{-3}$, the NS radius relaxes to a final value in $\order{10~\mathrm{ms}}$, well below the age of the youngest observed neutron star. This result indicates that for the physical times considered in our results, the calculated observables are insensitive to the value of the nucleonic friction coefficient $\alpha$.

\begin{figure}[t!]
    \centering
    \includegraphics[width=0.5\textwidth]{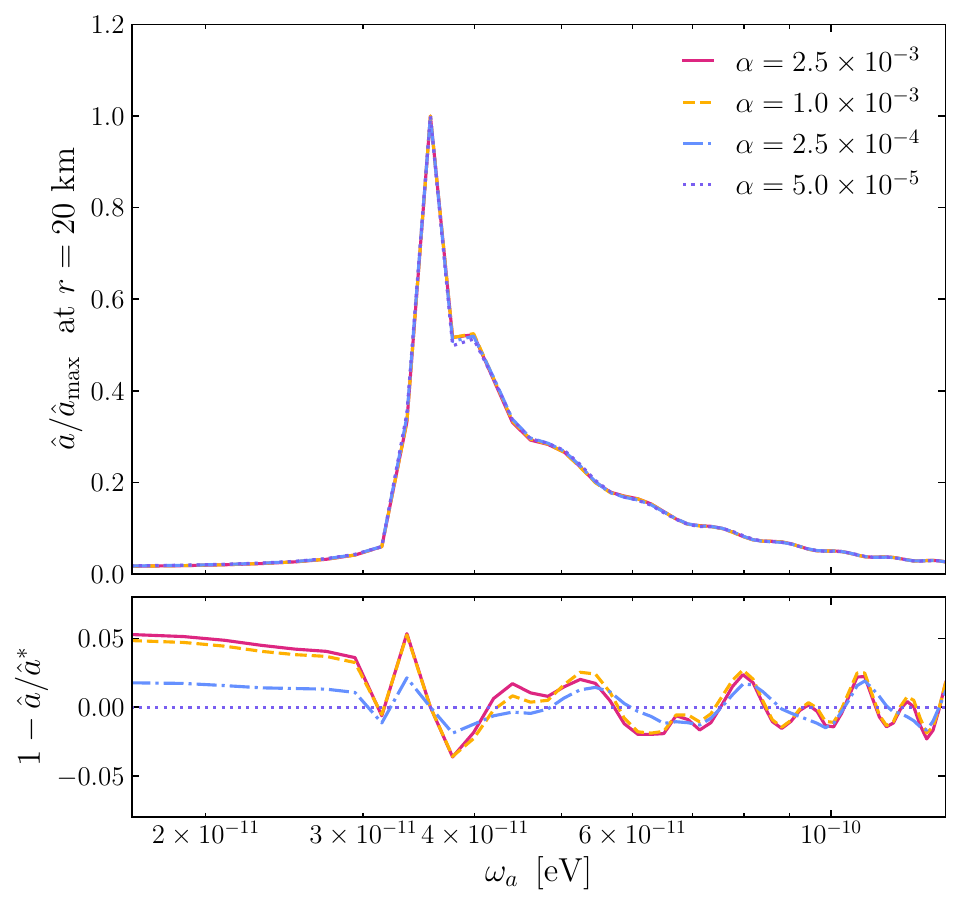} \\
    \caption{Axion spectrum at $r = 20~\mathrm{km}$ plotted for four different nucleon sector friction coefficients $\alpha$. The spectrum does not widen appreciably for larger values of the friction, indicating that the dominant source of damping for axion dynamics far outside the star is the axion-metric coupling.}
    \label{fig:frequencies-friction}
\end{figure}

It is also informative to understand how the decay of the axion dynamics outside the neutron star is affected by the added nucleonic damping in Eq.~(\ref{eq:NS-normal-damping-app}). In Fig.~\ref{fig:frequencies-friction}, we plot the normalized Fourier coefficients of the axion field at $20~\mathrm{km}$, well outside the radius of the NS. The resultant width of the axion Fourier coefficients is insensitive to the NS normal mode damping coefficient $\alpha$, indicating that the primary method of axion damping outside the star is through the axion-metric coupling.

\subsection{Initial Condition Dependence}
\begin{figure}[h!]
    \centering
    \includegraphics[width=1.0\textwidth]{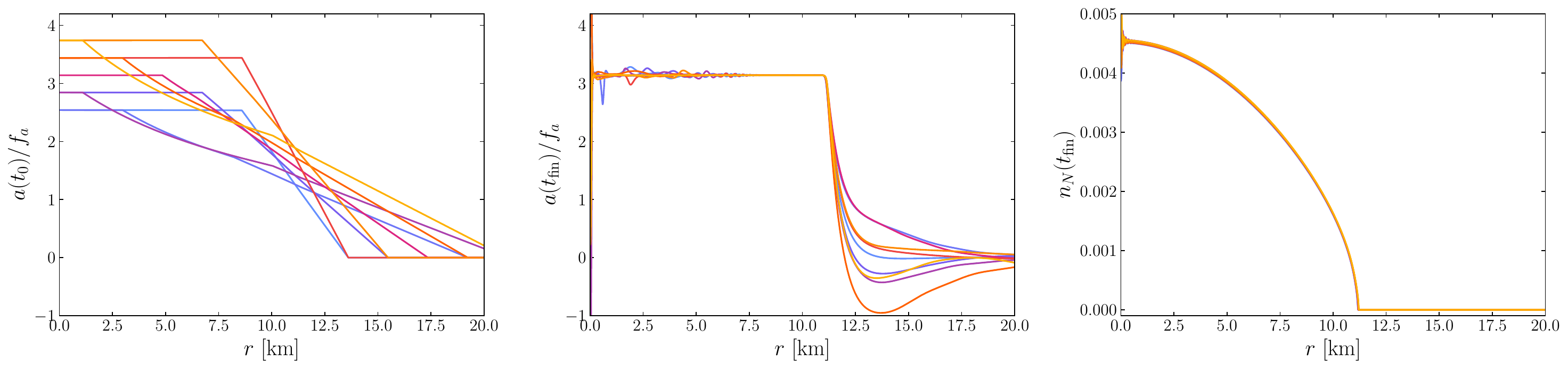} \\
    \caption{Test of the sensitivity of the final state to initial axion field profile. Initial axion profiles are shown in the left panel, final axion profiles are shown in the center panel, and final state nucleon profiles are shown in the right panel. The internal solutions for the axion are independent of the initial condition up to any undamped dynamics. The final state of the nucleon number density, which dominates NS observables, is independent of initial state.}
    \label{fig:ic-test}
\end{figure}

For the simulation results presented here, the initial condition of the axion field is defined by
\begin{equation}
    \label{eq:axion-initial-state-app}
    a(t = 0, r) = \pi f_a \Theta\left( r_{\rm crit} - r \right) \otimes \mathcal{W}
\end{equation}
where $\mathcal{W}$ is the window function $\mathcal{W}(r, w) = \Theta(r + w) - \Theta(r - w)$ and $r_{\rm crit}$ is defined implicitly via $n_N(r_{\rm crit}) = n_N^{\rm crit}$. In Fig.~\ref{fig:ic-test}, we show that the final results of the simulation are insensitive to the initial condition of the axion field. In particular, we show that the nucleon number density and the axion field on the interior of the star, which govern the observables we consider, are consistent across a wide range of initial axion profiles.

\subsection{Integration Step Dependence}
\begin{figure}[t!]
    \centering
    \includegraphics[width=1.0\textwidth]{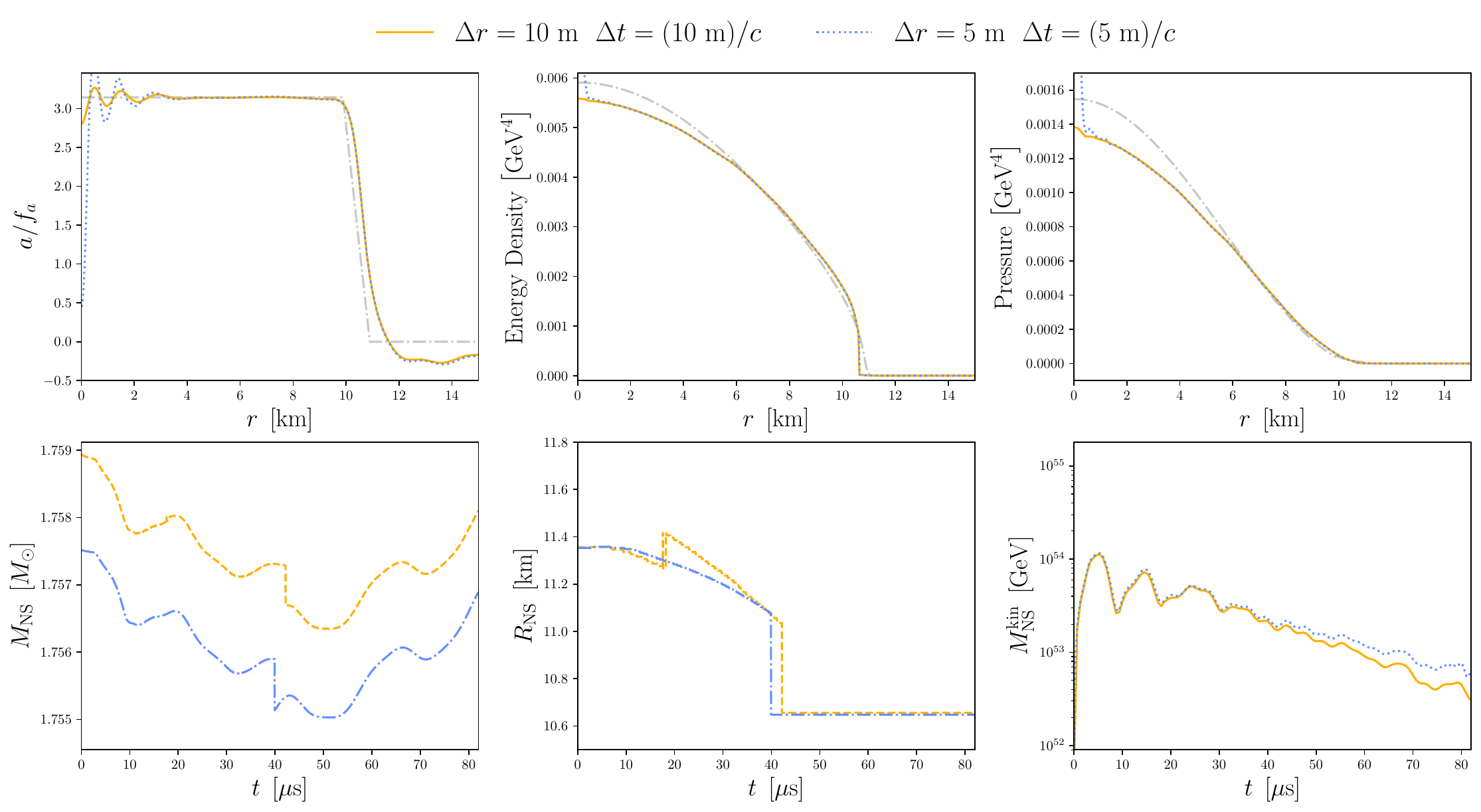} \\
    \caption{Test of the sensitivity of final state to varying integration steps $\Delta r$ and $\Delta t$. We simulate the fields with half the radial and temporal integrations steps used in the simulations presented here and produce consistent results. Initial state (dot-dashed, grey) axion field and final state fields for three sets of integration steps (TOP LEFT). Initial and final state total energy density (TOP CENTER) and total pressure (TOP RIGHT). Neutron star mass (BOTTOM LEFT), radius (BOTTOM CENTER), and total kinetic contribution to the NS mass (\ref{eq:kinetic-energy-density}) (BOTTOM RIGHT) as a function of physical time.}
    \label{fig:steps}
\end{figure}

To verify the stability of the simulations presented here, we perform simulations with smaller temporal and radial steps and show that the results are consistent. In Fig.~\ref{fig:steps}, curves for a representative simulation with initial central nucleon number density $n_N^c = 4.5~n_N^{\rm sat}$ and with integration steps $(\Delta r = 10~\mathrm{m}, \Delta t = 10~\mathrm{m} / c)$ and $(\Delta r = 5~\mathrm{m}, \Delta t = 5~\mathrm{m} / c)$ are shown, both of which exhibit the same behavior, validating our numerical integration methods.

\end{document}